\documentclass[fleqn,usenatbib]{mnras}
\usepackage{newtxtext,newtxmath}

\usepackage[T1]{fontenc}
\usepackage{ae,aecompl}

\usepackage{graphicx}	
\usepackage{amsmath}	
\usepackage{amssymb}	

\title[SuperNNova: SN Bayesian photometric classification]{SuperNNova: an open-source framework for Bayesian, Neural Network based supernova classification}

\author[M{\"o}ller \& de Boissi{\`e}re]{
A. M{\"o}ller,$^{1,2,3}$\thanks{anais.moller@clermont.in2p3.fr}
T. de Boissi{\`e}re$^{4}$ \thanks{Equal contribution}
\\
$^{1}$Research School of Astronomy and Astrophysics, Australian National University, Canberra, ACT 2611, Australia\\
$^{2}$ARC Centre of Excellence for All-sky Astrophysics (CAASTRO)\\
$^{3}$Universit{\'e} Clermont Auvergne, CNRS/IN2P3, LPC, F-63000 Clermont-Ferrand, France\\
$^{4}$Lyrebird AI, Unit 302, 55 Mont-Royal Avenue Ouest, Montreal, QC H2T 2S5, Canada.
}

\date{Accepted XXX. Received YYY; in original form ZZZ}

\pubyear{2018}

\begin{document}
\label{firstpage}
\pagerange{\pageref{firstpage}--\pageref{lastpage}}
\maketitle

\begin{abstract}
We introduce SuperNNova, an open source supernova photometric classification framework which leverages recent advances in deep neural networks. Our core algorithm is a recurrent neural network (RNN) that is trained to classify light-curves using photometric information only. Additional information such as host-galaxy redshift can be incorporated to improve performance. We evaluate our framework using realistic supernovae simulations that include survey detection. We show that our method, for the type Ia vs. non Ia supernovae classification problem, reaches accuracies greater than $96.92 \pm 0.09$ without any redshift information and up to $99.55 \pm 0.06$ when redshift, either photometric or spectroscopic, is available. Further, we show that our method attains unprecedented performance for classification of incomplete light-curves, reaching accuracies $>86.4 \pm 0.1$ ($>93.5 \pm 0.8$) without host-galaxy redshift (with redshift information) two days before maximum light. In contrast with previous methods, there is no need for time-consuming feature engineering and we show that our method scales to very large datasets with a modest computing budget. In addition, we investigate often neglected pitfalls of machine learning algorithms. We show that commonly used algorithms suffer from poor calibration and overconfidence on out-of-distribution samples when applied to supernovae data. We devise extensive tests to estimate the robustness of classifiers and cast the learning procedure under a Bayesian light, demonstrating a much better handling of uncertainties. We study the benefits of Bayesian RNNs for SN Ia cosmology.
Our code is open-sourced and available in github \footnotemark.
\end{abstract}


\footnotetext{\url{https://github.com/supernnova/SuperNNova}}



\section{Introduction}
To study the expansion of the Universe, increasingly large samples of Type Ia supernovae (SNe Ia) are used to measure luminosity distances as a function of redshift. Traditionally, SNe Ia are classified after spectroscopic follow-up of promising supernova candidates identified by photometric surveys. Previous and current surveys such as SNLS, SDSS, the Foundation Supernova Survey and Dark Energy Survey have discovered thousands of supernovae \citep{Astier:2005,Frieman:2008,Foley:2018,Bernstein:2011}. Of those supernovae, only a small number can be spectroscopically followed-up. The upcoming Large Synoptic Survey Telescope (LSST) survey is expected to discover $10^7$ supernovae \citep{LSST:2009}. This survey, together with others such as the Zwicky Transient Survey (ZTF), will explore transient events and expect to have up to a million transient alerts per night \citep{Bellm:2018}. The scarcity of spectroscopic resources is, and will continue to be, a limiting factor for supernovae cosmology and time-domain astronomy. Thus, in order to fully take advantage of the available data, fast and accurate photometric classification is crucial.

In recent years different methods have been developed to classify supernovae using their light-curves, i.e. their observed brightness evolution over time in different color bands. This is a challenging problem since brightness measurements are often noisy and non-uniform in time. Most classification methods rely either on light-curve template matching \citep{Poznanski:2007,Sako:2011} or classification using features extracted from template or functional fits. Extracted features can be used to classify supernovae using sequential cuts \citep{Bazin:2011,Campbell:2013} or machine learning algorithms \citep{Kessler:2010r,Moller:2016,Lochner:2016,Dai:2017}. Recent advances in Deep Learning, a branch of machine learning, have shown that neural network classifiers trained on raw data can outperform classifiers based on handcrafted features \citep{Charnock:2017,Kimura:2017,Moss:2018}. In this work, we explore this promising direction and obtain state-of-the-art results on a variety of supernovae classification tasks.

Typically, the performance of classification algorithms is assessed on simulated data. In particular, many algorithms have used the ``Supernova Photometric Classification Challenge'' (SPCC) \citep{Kessler:2010} dataset for training and evaluation. Some of these methods have also been successfully applied to real survey data, such as SDSS \citep{Sako:2011} and SNLS \citep{Kuznetsova:2007,Poznanski:2007,Bazin:2011,Moller:2016}. Recently, in preparation for LSST, an open data challenge called ``Photometric LSST Astronomical Time Series Classification Challenge (PLAsTiCC)" \citep{Plasticc_data:2018} was launched in an effort to better classify simulated astronomical time-series data.

Existing classification algorithms have been shown to obtain high-purity SN Ia samples with complete light-curve information. However, only a handful are able to classify SNe within a limited number of photometric epochs \citep{Poznanski:2007,Sako:2011,Ishida:2012,Charnock:2017,Kimura:2017}. Partial light-curve classification is key to prioritizing spectroscopic and photometric resources for promising candidates. 

Photometrically classified supernovae can be used in statistical analyses such as rates and SN Ia cosmology \citep{Gonzalez-Gaitan:2011,Campbell:2013,Jones:2018} but those analyses suffer from bias induced by contamination from other SN types. For type Ia SN cosmology, this bias has been accounted for using Bayesian frameworks \citep{Hlozek:2012,Jones:2018,Hinton:2018} but little attention has been devoted to the calibration of the underlying machine learning classifiers. This should be an important concern given that prior work has demonstrated that miscalibration was a common issue in machine learning \citep{Niculescu:2005,Guo:2017}. Another pitfall is that commonly used algorithms give no information about the uncertainty of their predictions. Recent advances in Bayesian Neural Networks \citep{Gal_a:2015,Gal_b:2015,Fortunato:2017} provide a framework for training neural networks that are able to estimate model uncertainty in a computationally efficient way.

Model uncertainties can provide valuable information on the completeness of the training set. It has been shown that the data samples commonly used to train supernovae classifiers are far from complete \citep{Lochner:2016,Ishida:2018}. These samples are further biased by the fact that they only contain known supernova types which match specific templates. New classes of transients may be found in the survey sample and existing classification algorithms will still try to assign it to one of the known classes or targets. Such events, so called out-of distribution events can contaminate a photometrically selected sample.  To avoid this, some analyses advocated for pre-selection cuts \citep{Kuznetsova:2007,Moller:2016} to restrict candidates to a subset of known classes. However, this prevents the analysis of new cosmological objects, a task which will be a major focus of brokers such as ANTARES \citep{Narayan:2018}.

In this work, we present SuperNNova, an open source algorithm that fits the requirements highlighted above and alleviates some of the concerns linked to the use of machine learning. SuperNNova is based on raw, photometric time-series inputs. Thus, it is not biased by template matching nor limited by costly feature engineering. 

SuperNNova is the first SN light-curve classification framework that uses Bayesian Neural Networks. For the first time in astronomical analyses, two Variational Inference methods \citep{Gal_b:2015,Fortunato:2017} are available in the same framework and compared using SNe simulations. Furthermore, we are able to handle irregular time-series without the use of data augmentation nor fitting methods. This is crucial for telescope observations which are subject to weather condition changes and non-uniform scheduling.

With SuperNNova, we obtain state-of-the-art results on the classification of supernovae with complete and partial light-curve data. We show that SuperNNova's performance increases with the amount of data and show that it readily scales to very large datasets. The performance is further enhanced with additional information such as host-galaxy redshifts. Crucially, we show that SuperNNova's classifiers can be trained in a principled, Bayesian way and yield calibrated predictions with sensible uncertainty estimates. Those uncertainty estimates make SuperNNova better suited to the task of identifying out-of-distribution candidates as we show it is not susceptible to classifying those samples with high confidence.

Although this method has been designed with supernova cosmology in mind, SuperNNova can also be applied to any light-curve classification task in time-domain astronomy.

This manuscript is structured as follows: we will introduce the simulations, metrics and machine learning algorithms used in Section~\ref{section:methods}. Then we will present and study the performance of the three main algorithms available in SuperNNova: a Baseline RNN (Section~\ref{section:BaselineRNN}) and two Bayesian RNNs (Section~\ref{section:BayesianRNNs}). In Section~\ref{section:resources} we briefly outline the computational resources required to use SuperNNova. In Section~\ref{section:towardscosmology}, we devise extensive tests to evaluate the statistical robustness of our classifiers and analyze the effect of a photometrically classified type Ia SN sample in cosmology.

\section{Methods}\label{section:methods}

  \subsection{Simulations and available datasets} \label{section:simulations}
  We simulated realistic supernovae light-curves using the SNANA package \citep{Kessler:SNANA}. Our data is similar to the SPCC data \citep{Kessler:2010,Kessler:2010r} with updated models used in the DES simulations \citep{Kessler:2018}. DES light-curves are built from supernova templates and use SALT2 SN Ia SED models \citep{Guy:2007} and the trained model from JLA \citep{Betoule:2014}. Observing logs specifying the simulated cadence and conditions were included when available.
  
  Our simulation, which is publically available \footnote{\href{https://zenodo.org/record/3265189}{DOI:10.5281/zenodo.3265189}},  contains $1,983,213$ supernovae light-curves of which $881,969$ are successfully fitted using SALT2 to obtain features such as color and stretch. SALT2 fits do not always converge and therefore some light-curves can not be fitted. Furthermore, we require loose cuts for fitting, such as: an observation with $S/N >5$, and at least 1 epoch before $-2$ days, at least $1$ epoch past $+30$ days (rest-frame). The simulation includes spectroscopic and photometric host galaxy redshifts \citep{Kessler:2010,Gupta:2016}. To reflect real survey conditions, spectroscopic redshift is obtained only for a subset of light-curves. 
  
    In all the classification tasks that follow, we sub-sample the dataset to make sure classes are balanced. For binary classification i.e. the discrimination of SNe Ia vs. others, we obtain a balanced sample of $912,691$ light-curves (resp. $402,786$ with SALT2 fits). Detailed statistics for binary classification are shown in Table~\ref{table:simulation-types}. Our simulation count is more than an order of magnitude larger than SPCC which contained $21,319$ light-curves. We split the data as follows: 80\% training, 10\% validation and 10\% testing.
    
    In the following we report our results separately for the {\it ``complete''} and the {\it ``SALT2 fitted''} datasets. The former contains all our simulated light-curves after class-balancing. The latter contains only light-curves that were successfully fitted with SALT2 \citep{Guy:2007}. Since past and current supernova surveys have been focused on discovering type Ia supernovae, this {\it ``SALT2 fitted''} sample is closer to their spectroscopically targeted supernovae. We will call the {\it ``SALT2 fitted''} dataset a non-representative one, as it fails to explore the full diversity of supernovae. Our representative set will be the {\it ``complete''} dataset. We further explore these samples and the issue of representativeness in Section~\ref{section:representative}.

    \begin{table}
    \centering
    \caption{Simulated supernovae samples divided by subtypes. We show the number of supernovae light-curves available per type in the complete dataset and the number that have a succesful SALT2 fit. Each SN type represents a template with the exception of type IIL for which we use two different templates.}
      \label{table:simulation-types}
  \begin{tabular}{l  cr } 
    \multicolumn{3}{c}{Simulated supernovae} \\
        \hline
    SN type & SALT2 fitted & complete dataset \\
    \hline
    Ia   &             402,786 &                912,691 \\
    Ib   &             140,197 &                181,454 \\
    Ic   &              70,811 &                 90,485 \\
    IIP  &              94,994 &                296,523 \\
    IIn  &               3,249 &                154,614 \\
    IIL &               93,535 &                189,615 \\
    \hline
    \multicolumn{3}{c}{IILs by template} \\
    \hline
    IIL1 &              26,717 &                100,827 \\
    IIL2 &              66,818 &                 88,788 \\
    \hline
  \end{tabular}
\end{table}

    \subsection{Evaluation metrics}
     \subsubsection*{ROC curve}
AUC is a robust metric, commonly used as an evaluation method for binary classification. AUC stands for Area Under Curve, where the curve is the ROC curve (Receiver Operating Characteristic). The ROC curve gives an indication of the performance of a binary classifier by plotting the true positive rate (efficiency) against the false positive rate (contamination).
  While the ROC curve represents the performance of a model in two-dimensions, the AUC simplifies this into a number. A perfect model would score an AUC of 1 while a random classifier would score $0.5$.

  \subsubsection*{Accuracy, Purity and Contamination}
  If classes are balanced, classification accuracy is an intuitive and useful metric. Accuracy is measured as the number of correct predictions against the total number of predictions.
  For binary classification, accuracy can also be calculated in terms of positives and negatives as follows:
  \begin{equation}
  \mathrm{accuracy} = \frac{\mathrm{TP+TN}}{\mathrm{TP+TN+FP+FN}}
  \end{equation}
  where TP (resp. TN) are true positives (resp. negatives) and FP (resp. FN) are false positives (resp. negatives). For binary classification, TP shows the number of correctly classified SNe Ia while TN shows the number of correctly classified core-collapse SNe. In this work, unless specified, the predicted type corresponds to the highest probability class. 

  The purity of the SN Ia sample and the classification efficiency are defined as:
  \begin{equation}
  \mathrm{purity} = \frac{\mathrm{TP}}{\mathrm{TP+FP}}; \quad \mathrm{efficiency} = \frac{\mathrm{TP}}{\mathrm{TP+FN}}
  \end{equation}\label{eq:purity}

  Contamination by core collapse is defined as:
\begin{equation}
\mathrm{contamination} =\frac{\mathrm{FP}}{\mathrm{FP + TN}}\label{eq:false_positive}
\end{equation}

In this work, performance metrics reported with errors represent the mean and one standard deviation of the distribution. This distribution is obtained by performing 5 runs with different random seeds.

  \subsection{Recurrent Neural Networks (RNNs)}
    RNNs are a class of neural nets that allow connections between hidden units with a time delay; they are well-suited for modeling sequential data or time series data \citep{Bahdanau_2014,Sutskever:2014,Mehri_2016, Kalchbrenner_2018, Vasquez_2019}. Through the connections between hidden units, the model can learn to retain or discard past information about its inputs, and in principle discover correlations across time. Popular recent architectures such as Long short-term memory (LSTM) \citep{Hochreiter:1997} or Gated Recurrent Univt (GRU) \citep{Chung:2014} alleviate some of the issues that occur when training these models. They introduce a memory cell and gating mechanisms that endow recurrent networks with better control of the information flow.

\begin{figure}
	\includegraphics[width=\columnwidth]{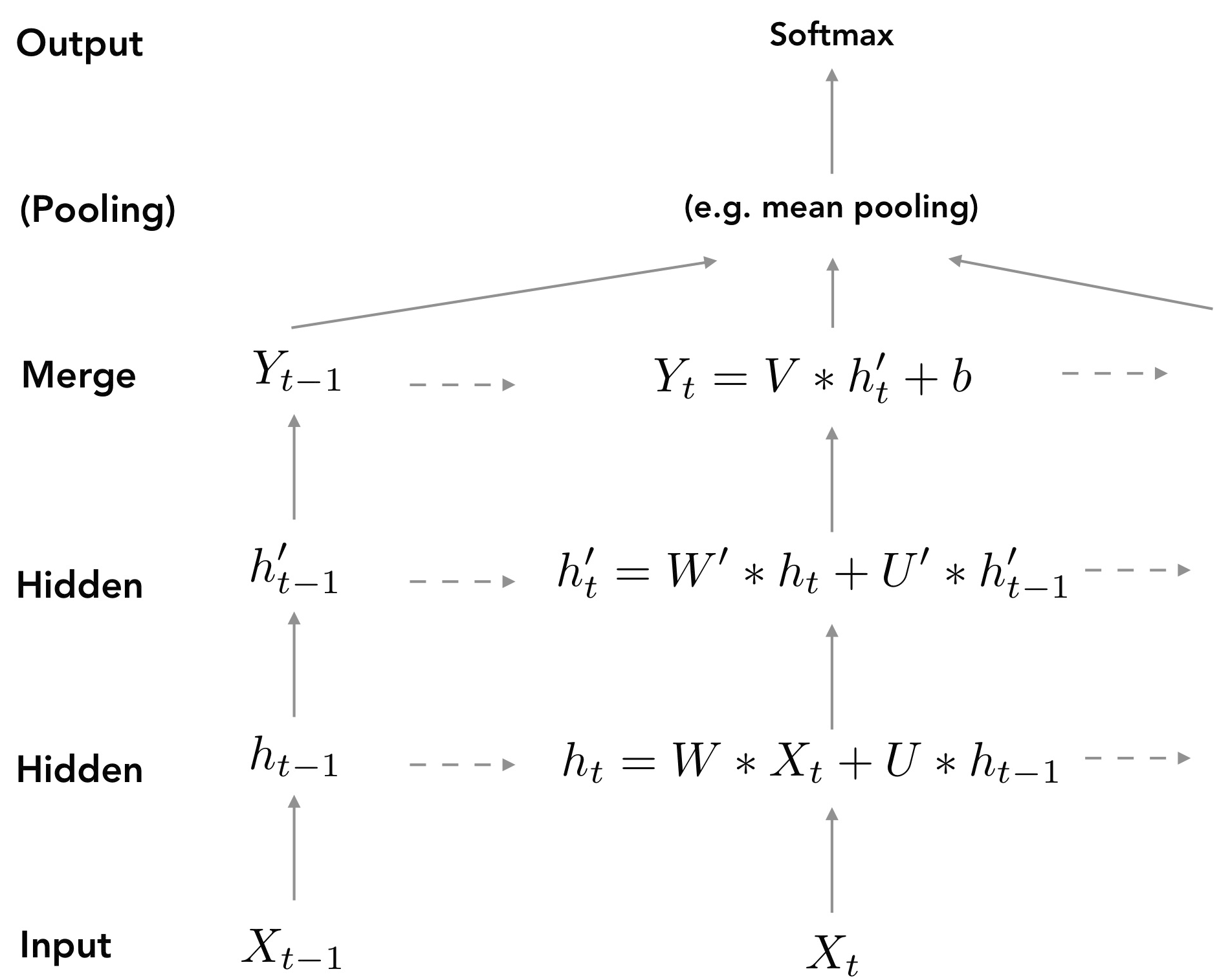}
    \caption{Partial schema of a vanilla RNN with mean pooling. Layers are indicated in the first left column. Inputs ($X_t$) are given to the network. Arrows indicate transmission of temporal information. The network extends to the right by the number of $t \in [1,T]$ available inputs.}
    \label{fig:RNN_schema}
\end{figure}

    The structure of a basic RNN with 2 hidden layers is illustrated in Figure~\ref{fig:RNN_schema}. The same series of operations (typically an affine matrix multiplication followed by an activation function) are applied to each element of a sequence ${(X)_{t \in [1,T]}}$ to produce a sequence of hidden states (${h_t}$). These operations are parameterized by learnt weights ($W,W',V,U,U'$ in Figure~\ref{fig:RNN_schema}) and biases ($b$). Multiple recurrent layers can be stacked on top of each other by feeding a layer's output as input to the next layer. To obtain the class prediction, the sequential information is eventually condensed to a fixed length representation (for instance through mean-pooling, i.e averaging of the hidden states). The use of a softmax function ensures that the network's output can be understood as probabilities. In Figure~\ref{fig:RNN_schema}, the RNN is unidirectional: it is allowed to process information from left to right,  bidirectional networks are also allowed to process from right to left. This has been used in many applications such as language translation to improve performance.

    Traditional RNNs lack the ability to quantify model uncertainty. However, quantifying the uncertainty of a model's prediction is a highly desirable feature for cosmological applications. In this work, we leverage two recent advances in variational inference that have proven effective in helping neural networks to estimate uncertainty. Gal et al. \citep{Gal_a:2015} show that dropout (a regularization technique consisting in randomly dropping weights or activations) has a Bayesian interpretation and can be used to improve performance and calibrate predictions. Fortunato et al. \citep{Fortunato:2017} optimize a posterior distribution over each of the neural network's weights and have shown competitive results and high quality uncertainties on challenging language modeling and image captioning tasks. We introduce variational inference and our approximation methods in Section~\ref{section:BayesianRNNs}.

    \subsection{RNN architecture} \label{section:RNN_architecture}

    Our PyTorch \citep{paszke2017automatic} implementation is publicly available on GitHub \footnote{\url{https://github.com/supernnova/SuperNNova}} with accompanying documentation.

    The core building block of our algorithm is a Recurrent Neural Network (RNN). Our RNN processes the photometric time series to produce a sequence of hidden states. The sequence is condensed to fixed length through mean pooling or alternatively by concatenating the last hidden state of each recurrent layer. In the following, we will call these RNN output options mean pooling or standard respectively. Finally, a linear projection layer is applied to obtain an $N$-dimensional vector, where $N$ is the number of classes. A softmax function is used to obtain probabilities for an input to belong to a given class.

    We make it easy for the user to experiment with hyper-parameters such as layer type, network size, bidirectionality, batch size or gradient descent optimizer through an extensive command line interface. In addition, we implement two Bayesian training methods. The first one, dubbed ``MC Dropout'' implements \citep{Gal_a:2015} and the second one, which we simply call ``BBB'' implements \citep{Fortunato:2017} \footnote{in the github paper branch these methods are named MC dropout and Bayesian respectively.}. These two Bayesian methods provide principled regularization for the training of neural networks, and have been shown to provide superior uncertainty estimation, a crucial property for astronomy and cosmology applications.
    
    This framework uses the python libraries: pandas, numpy, astropy, matplotlib, pytorch, scikit-learn, seaborn \citep{pandas:2010,numpy:2011,astropy:2018,astropy:2013,Hunter:2007,paszke2017automatic,scikit-learn:2011,seaborn:2018}.

    \subsubsection*{RNN training data}

    The only inputs required by our algorithm are photometric time series or light-curves. However, any other information available can and should be used to improve the performance. Thus, we made it straightforward to add additional inputs, such as host-galaxy redshift by simply repeating the redshift value and its error for all time-steps in the time series. There is no need for tedious, time-consuming feature engineering or domain-expert knowledge. This is a decisive advantage of our method since it has been shown \citep{Lochner:2016} that the choice of the feature extraction method can significantly impact classification results.

    This data can be used to train RNNs for three distinct classifications tasks: binary (SNe Ia vs. non Ia), ternary (Ia, II, Ibc) and seven-way classification  (Ia, IIP, IIn, IIL1, IIL2, Ib and Ic).

    \subsubsection*{RNN training details}

    The light-curve data is typically formatted in a tabular way, where each row corresponds to one observation, in one specific filter. We group observations within eight hours of each other on a single row. Which is equivalent to grouping all the observed fluxes in a given night. If a given filter is not observed, we assign it a special value to indicate that it is missing. We add a delta time feature to indicate how much time has elapsed since the last observation. We also add a feature indicating which filter is available.
    
    The features used in the algorithm for a given time step are: fluxes (FLUXCAL in filters $g,i,r,z$), flux errors (FLUXCALERR in filters $g,i,r,z$), time step (delta\_time) and one-hot encoding representing available flux information (g, gi, gir, girz, giz, gr, grz, gz, i, ir, irz, iz, r, rz, z). Additional information can be provided such as host-galaxy redshift, spectroscopic or photometric, and its error.

    Light-curve fluxes and errors exhibit large variations which is why we paid special care to input data normalization. Three options are supported: no normalization, individual, ``per-filter" normalization and ``global" normalization. The latter two are obtained as follows: features (f) are first log transformed and then scaled. The log transform ($f_{\textrm l}$) uses the minimum value of the feature $min(f)$ and a constant ($\epsilon$) to center the distribution in zero as follows: $f_{\textrm l} = \log\left(-min(f) + f + \epsilon \right)$. Using the mean and standard deviation of the log transform ($\mu,\sigma$($f_{\textrm l})$), standard scaling is applied: $\hat{f} =  (f_{\textrm l} - \mu(f_{\textrm l})) / \sigma(f_{\textrm l}) $. In the ``global" scheme, the minimum, mean and standard deviation are computed over all fluxes (resp. all errors). In the ``per-filter" scheme, they are computed for each filter.

    We use the Adam optimizer \citep{Kingma:2014} to train our RNNs. Two learning rate policies were implemented: a standard ``decay on plateau" policy which decays the learning rate by a constant factor when the validation loss stagnates and the 1-cycle policy introduced in \citep{Smith:2018}.

    Dropout is applied to all our models for regularization (except in the ``BBB" case where regularization is enforced by a Kullback Leibler divergence term in the loss function). Owing to the large size of the dataset and limited overfitting observed, we did not use data augmentation. However, to improve performance with incomplete light curves, we found it beneficial to train the network with randomly truncated light-curves. To make better use of parallelism, we batch together multiple light-curves at a time, using zero-padding and masking where appropriate.

    \subsection{Random Forest (RF)} \label{section:RF}
    To compare our Recurrent Neural Networks with other classification methods, we implement an additional classifier. We use the excellent scikit-learn \citep{scikit-learn:2011} library to implement a random forest classifier, which has been previously shown to have high performance in supernova photometric classification tasks \citep{Moller:2016,Lochner:2016,Dai:2017}. Features for this classifier are extracted from complete light-curves with the SALT2 fitter \citep{Guy:2007}. This method has been shown to have good performance in binary classification with complete light-curves \citep{Lochner:2016}. We emphasize that the SALT2 fitter requires a redshift as input. Thus, even if no explicit redshift information is given to the random forest classifier, the information implicitly percolates through the SALT2 fit parameters. We retained the following features to train our random forest classifier: stretch ($x1,x1_{ERR}$), color ($c,c_{ERR}$), magnitude in B-band ($mB,mB_{ERR}$), normalization ($x_0,x_{0_{ERR}}$), goodness of fit ($\chi^2$), and observed magnitudes in $rigz$ bands ($m_{0obs_r},m_{0obs_i},m_{0obs_g},m_{0obs_z}$ and their errors).
    
    \subsection{Convolutional Neural Networks (CNNs)} \label{section:CNN}
    CNNs are another type of deep learning networks. They were originally introduced for image processing \citep{Lecun:1998}. CNNs are made up of a cascade of convolutional layers which perform sliding cross-correlation of learned weight kernels with the input data. In astronomy, they have been successfully applied to various problems including classification and discovery of astrophysical objects \citep{Kimura:2017,Gieseke:2017, Lanusse:2018,Carrasco:2018}.
    
    For comparison with our RNN architecture introduced in Section~\ref{section:RNN_architecture}, we implemented a simple CNN architecture in PyTorch \citep{paszke2017automatic}. Our network is made up of four 1D convolutional layers. Following~\ref{section:RNN_architecture}, we collapse the sequence to a fixed length representation and obtain the final prediction by applying a linear projection layer.

\section{Baseline RNN} \label{section:BaselineRNN}
\begin{figure}
\includegraphics[width=1\columnwidth]{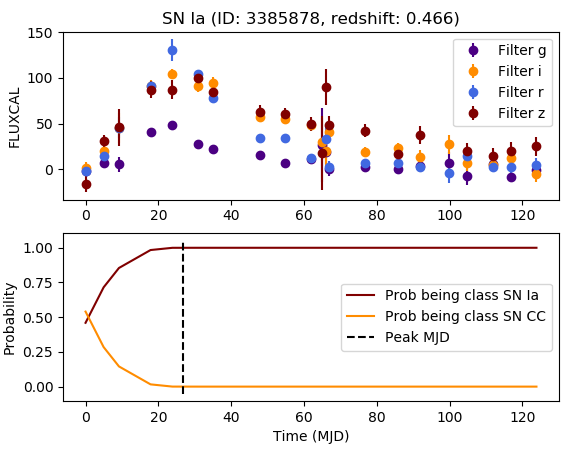}
\caption{Simulated type Ia supernova light-curve and classification. Top: calibrated flux evolution in different DES band-passes as a function of normalized time (the first photometric measurement is set to time equals zero). Bottom: Baseline RNN classification probability evolution with respect of time, no host-galaxy redshift information was provided. At each photometric measurement, classification probability is obtained. Maximum light of the simulated supernova is shown in a gray dashed line and simulated redshift of the supernovae is shown on the top $z = 0.466$. We highlight that redshift is  not used for this classification but can improve results. Our Baseline RNN classifies this light-curve as type Ia SN with great accuracy before maximum light, it only requires a handful of photometric epochs.}
\label{figure:baseline_lc}
\end{figure}

 We introduce a "Baseline RNN'' which is trained in the standard, non-Bayesian way. In this Section, we demonstrate the capabilities of this baseline model. We show that it is able to accurately classify light-curves at each time step as shown in Figure~\ref{figure:baseline_lc}. We study the impact on the performance when additional information such as host-galaxy redshift is added, when different training sets are used and when varying the number of classification targets (binomial, ternary and seven-way classification).
 
 \subsection{Hyper-parameters}
 As shown in Section~\ref{section:RNN_architecture}, different network architectures can be easily implemented in SuperNNova. We assess the impact of these variations using two metrics: the complete light-curve classification accuracy and the AUC for the binary classification task for $0.2$ of the {\it SALT2 fitted} dataset. We probed: batch size ($64, 128, 512, 1024$), optimizer (Adam, RMSProp), dropout on recurrent layers ($0.05, 0.1, 0.2$), layer type (GRU, LSTM), bidirectional (True, False), number of layers ($1,2$), rnn output (standard, mean pooling), random length (True, False) and the two learning policies. We also compared two ways to compress the hidden state sequence before the final prediction: last hidden state concatenation and mean pooling.
These hyper parameters do not have a strong impact on the performance of the Baseline RNN (classification accuracy for complete light-curve classification has a mean of $94.48$ and a standard deviation of  $1\%$).  
We obtain two sets of hyper-parameters with the same accuracy of $96.01$ without redshift information. We choose as our best hyper-parameters: batch size $128$, Adam optimizer, $0.05$ dropout on recurrent layers, bidirectional LSTM layers, $2$ recurrent layers, cyclic learning rate and standard rnn output. The latter however, is not preferred for partial light-curve classification which will be explored in Section~\ref{sec:partial-classification}. To improve early light-curve classification we choose the option of mean pooling before final prediction and training with randomly truncated light-curves. These settings will be used throughout the following unless specified.

\subsection{Normalization}
Input feature normalization is common practice to facilitate the training of neural networks. With half of the {\it SALT2 fitted} dataset, we evaluate the impact of different normalization types for fluxes and errors: no normalization, a global normalization over all filters and a per filter normalization. We find that the accuracy of our classifier is slightly improved when using global normalization ($96.12 \pm 0.07$) when compared to no normalization ($94.2 \pm 0.2$). We hypothesize that our use of optimizers with adaptive learning rate greatly mitigates feature scale issues. A per filter normalization is found to have equivalent performance ($96.0 \pm 0.2$) to global normalization. Such a normalization scheme blurs the color ratio between observed light-curves in different band-passes. Interestingly, this indicates that color ratios are not deemed an important feature for our algorithm's accurate classification of this supernova sample. An in depth study of normalization impact is out of the scope of this paper. Unless specified otherwise, we choose to apply global normalization throughout the rest of this work.

\subsection{Comparison with other methods}
The primary driver of supernova photometric classification has been to disentangle type Ia from other types of supernovae. We evaluate the performance of our SuperNNova Baseline RNN on this task and compare it with different classification algorithms: the Random Forest introduced in Section~\ref{section:RF}, the Convolutional Neural Network structure in Section~\ref{section:CNN} and results obtained with another RNN architecture \citep{Charnock:2017}. We use the same training and validation samples from the {\it SALT2 fitted} dataset to evaluate all algorithms. We assess uncertainties in the accuracy by performing 5 randomized runs that select different training, validation and testing splits. We obtain for this dataset an accuracy for the Baseline RNN of $96.3 \pm 0.4$ and an AUC of  $0.995 \pm 0.001$.

\subsubsection{Baseline RNN vs. CNN}
As we did for our baseline RNN, we first choose the best hyper-parameters for the CNN. We probed: batch size (64,128,512), random length (True, False), size of hidden dimension (16, 32, 64) and different learning rates (0.01,0.001,0.0001). Our best performance is found with: batch size of 128, hidden dimension 32, learning rate 0.01, bidirectional layers and training with random length sequences. For this configuration, we find that our CNN architecture obtains an accuracy of $94.53 \pm 0.07$ and an AUC $ 0.9887 \pm 0.0002$ without any redshift information for the whole {\it SALT2 fitted dataset}. Our Baseline RNN obtains higher accuracies for this classification task. This behaviour is also seen with a smaller data fraction, $0.2$ where classification accuracies are found to be $93.1 \pm 0.2$ ($97.1 \pm 0.1$, $98.98 \pm 0.07$) without (with photometric, spectroscopic) host-redshift information.
We note that other CNN configurations may yield better accuracies. Our CNN model is available in the SuperNNova repository\footnote{\url{https://github.com/supernnova/SuperNNova}} in order to foster further experiments on this classification task.

\subsubsection{Baseline RNN vs. RF}
The RF classifier obtains accuracies of  $95.15 \pm 0.07$ and an AUC of $0.9929 \pm 0.0003$ without redshift information. Our Baseline RNN has a higher accuracy and AUC than many RF methods found in the literature \citep{Moller:2016,Lochner:2016,Dai:2017}. However, classification accuracies depend on both the size and content of training and testing sets and therefore cannot be directly compared. We study the effect of the size of the training sample on the accuracy. As shown in Figure~\ref{fig:accuracy_methods}, we find that our Baseline RNN outperforms the Random Forest for large dataset sizes given the same redshift information. We also find that RNNs exhibit an increase in performance with larger datasets while the Random Forest classifiers remains almost constant. Again, we stress that even when no explicit redshift information is provided to the Random Forest classifier, the features obtained with the SALT2 fit implicitly contain this information. The high-accuracy of our RNN shows that it has successfully learned meaningful representations on top of raw photometric time-series, rather than using handcrafted features extracted from light-curve fitting. By using photometry directly we avoid biasing our classified sample, a particularly relevant observation since the choice of feature extraction has a strong impact on performance \citep{Lochner:2016}.

\subsubsection{Baseline RNN vs. another RNN architecture}
 For reference purposes, we compare a recently introduced RNN-based SN classifier with our Baseline RNN trained with a similar size set. \cite{Charnock:2017} uses half of the PCC dataset 5 times data-augmented, equivalent to $\approx 50,000$ supernovae light-curves, and obtains an accuracy of $93.1$ on binary classification without redshift information and $94.8$ with redshift. By selecting a data fraction of $0.05$ of our dataset, we can train our baseline RNN with $32,222$ light-curves and obtain a classification accuracy of $ 92.4 \pm 0.9$ and an AUC of $ 0.980 \pm  0.004$ without redshift information, when using photometric redshifts we obtain an accuracy of $ 96.8 \pm 0.1$. Although both methods were trained with different samples, our accuracies are comparable. Our algorithm seems to be more sensitive to redshift information providing up to $4\%$ accuracy increase for this sample size.

 \subsection{Redshift, contamination and efficiency}
 The addition of photometric redshifts, which are available for all our simulated supernovae, increases the accuracy of our baseline RNN by $2\%$ for photometric and $3\%$ for spectroscopic redshifts. Since spectroscopic redshifts are available for a subset of supernovae, we only evaluate performance on light-curves for which this redshift is available. We will continue using this subsample performance throughout the rest of this manuscript.

A photometrically classified type Ia supernova sample is expected to have a small percentage of contamination from other supernova types. As mentioned in Section~\ref{section:simulations}, our simulations are realistic and therefore include known SN rates and detection efficiencies for a survey such as DES. Therefore, our estimates are a good indicator of the expected contamination by core-collapse SNe in a photometrically classified SN Ia sample. 

We find that the contamination (or False Positive Rate) is dominated by types Ib and Ic supernovae when classifying without redshift information. This contamination peaks at simulated redshift between $0.2$ for Ic and $0.4$ for Ib and for our Baseline RNN is in the order of a couple percent for Ib and Ic SNe and under $1\%$ for other types. Contamination levels are found to be not correlated with the number of light-curves for training. Within this $1\%$ contamination there are type IIL and IIP SNe which are more numerous than type Ic in our simulations but are mostly correctly classified by SuperNNova. Host-galaxy redshifts increase the performance of our classifier reducing contamination. In particular simulated supernovae from Ib, and Ic  templates are better classified, reducing their contamination contribution in our Baseline RNN from  $2.5 \pm 0.2$ and $1.3 \pm 0.1$ respectively to $<0.1\%$ with host redshifts. Interestingly, classification of other core-collapse SNe such as IIP, IIn and IIL1 is barely affected by this additional information.

Selection efficiency is an important metric when classifying type Ia supernovae. For spectroscopically selected samples, selection efficiencies drop quickly for faint events \citep{Kessler:2018,Dandrea:2018}. Our Baseline RNN performs superbly, with almost constant efficiency as a function of simulated redshift with an average efficiency of $97.3 \pm  0.4$ without and up to $99.61 \pm 0.09$ with host-galaxy redshift information. Such high efficiencies enable probing new supernovae populations at high-redshift and can have an important effect on selection biases found currently in cosmology as will be discussed in Section~\ref{section:towardscosmology}.

Furthermore, one of the biggest limitation of photometrically classified samples is the expected contamination level by other supernova types which may affect statistical analyses \citep{Hlozek:2012,Jones:2018}. It has also been shown that for photometrically classified SNe Ia there is a compromise between sample purity and the efficiency of the classifier \citep{Moller:2016,Dai:2017}. We find that our highly efficient algorithm does not compromises the purity of the sample. On the contrary, the purity of a SN Ia sample classified without redshift is $ 95.4 \pm 0.5$, while the addition of redshifts increases the sample purity to $99.49 \pm 0.05 $. This level of contamination, less than $1\%$, is within the current range of contamination of spectroscopically classified samples currently used in cosmological and astrophysical analyses \citep{Rubin:2015}. 

\begin{figure}
\includegraphics[width=\columnwidth]{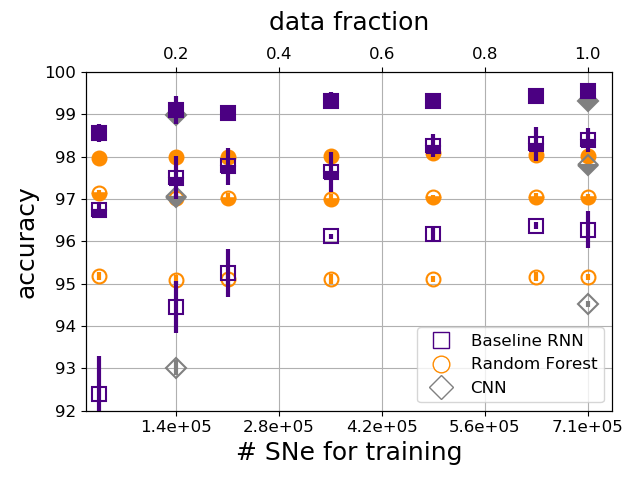}
\caption{Ia vs. non-Ia classification accuracy with respect to training set size using the {\it SALT2 fitted} dataset. Top axis indicates the used data fraction and the bottom axis number of supernovae. Accuracy of baseline RNN is shown in indigo squares, CNN in grey diamonds and Random Forest in orange circles. Error bars are one standard deviation from the accuracy distribution for five runs with different random seeds to initialize the networks weights. Classification using: no host-galaxy redshifts are empty, photometric host-galaxy redshifts bottom filled and spectroscopic host redshifts fully filled. Our Baseline RNN achieves greater accuracies than the other classifiers for the same dataset and explicit redshift information.}
\label{fig:accuracy_methods}
\end{figure}

\subsection{Early light-curve classification} \label{sec:partial-classification}
Having demonstrated the high performance of our baseline RNN in comparison to other classification methods, we now turn to explore other capabilities of SuperNNova. To do so, we use the {\it complete} dataset which contains light-curves that are not successfully fitted with SALT2. This increases the number and diversity of the training set for both core-collapse and type Ia supernovae. The effect of the SN diversity on the classifier performance is further studied in Section~\ref{section:representative}.

While there are currently a wealth of algorithms that can classify complete light-curves, only a handful are able to classify partial light-curves. Our RNN architecture allows us to accurately classify supernovae with a limited number of photometric epochs. At each light-curve time step, as more information is added, the RNN hidden state is updated, allowing the network to adapt its predictions.

\begin{figure}
	\includegraphics[width=\columnwidth]{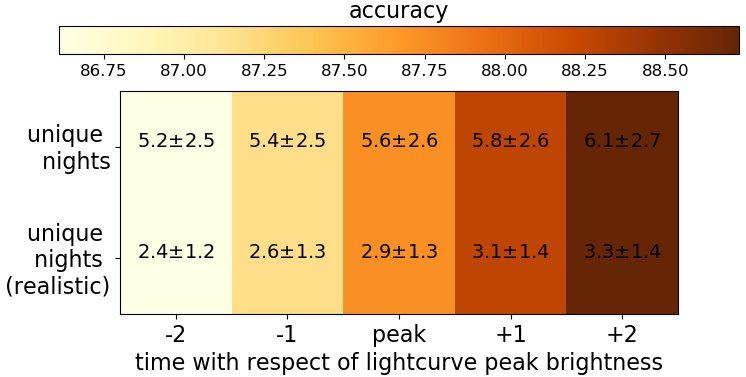}
    \caption{Average number of photometric epochs (nights) used for early light-curve classification. Rows indicate two ways of computing the average number of photometric nights available for classification while columns indicate the days around simulated peak brightness of the supernova (-2,-1, at peak, +1,+2). The raw average of photometric epochs is shown in the row "unique nights", since our simulated light-curves can include observations before the supernova explosion, a more realistic estimation of the useful epochs include only observations two weeks before observed peak brightness which is a loose estimate of the pre-maximum phase of a type Ia supernova light-curve, this estimation is shown in the row "unique nights (realistic)". Colors indicate the accuracy obtained for these classifications with the Baseline RNN as shown in  Table~\ref{table:partial-lcs} for the {\it complete} dataset without any redshift information.
    }
    \label{figure:n_nights}
\end{figure}

In Table~\ref{table:partial-lcs} we present our baseline RNN accuracies for partial light-curve classification for the whole {\it SALT2 fitted} and {\it complete} datasets. The number of photometric epochs (nights) available at different stages of the light-curves are shown in Figure~\ref{figure:n_nights}. In average, we require only $\approx 5$ observing nights to obtain classification accuracies $>80\%$. Since simulated light-curves can contain photometric nights before the actual supernova explosion, we provide a "realistic" estimate for unique nights useful in the classification. For this estimate, we select only photometric epochs within 14 days before peak maximum which is a loose threshold for the rising time of type Ia supernovae. Using this more realistic metric, we require between $2.4 \pm 1.2$ and $3.3 \pm 1.4 $ photometric epochs in average to start accurately classifying supernovae. We note that this low number of required epochs is linked to multi-color observations in each night which provide valuable information for classification.

Additionally, since training set numbers affect performance metrics, we compare the performance with $43\%$ of the {\it complete} dataset. This fraction of the complete dataset has a similar size training set than the {\it SALT2 fitted} one.

For all datasets, our baseline RNN is capable of attaining a high-accuracy classification, $>83\%$ right before the maximum of a supernova light-curve. Further, we show that the addition of host-galaxy redshifts produce a rise in accuracy as high as $12$ points. 

We note that for spectroscopic redshifts, the {\it complete} sample has lower accuracy when compared with the use of photometric redshifts. This was not observed in the {\it SALT2 fitted} sample and it may be explained by a selection bias in the {\it complete} dataset for supernovae with spectroscopic redshifts. 

When classifying type Ia supernovae before or around maximum light, we find that contamination is still dominated by Ib supernovae with $9.4 \pm 0.3 \%$ contribution and Ic with $3.3 \pm 0.2\%$. Interestingly, type IIP and IIL2 supernovae can contribute around $2\%$ of the contamination each, while this is rarely the case for complete light curve classification. This may be due to characteristic features of these light-curves only present after maximum light, such as the plateau exhibited by type IIp SNe.

In summary, we have shown that SuperNNova is able to accurately classify light-curves before and at maximum light. With accuracies ranging from $83.6 \pm 0.6$ up to $97.0 \pm 0.2$ for the {\it salt fitted} dataset, without and with redshifts respectively. SuperNNova opens a path towards efficient use of photometric and spectroscopic resources for follow-up. Candidates can then be prioritized for diverse science goals including targeted samples (e.g. SNe Ia for cosmology) and improving the SN sample for photometric classification as recently proposed by \cite{Ishida:2018}. Such a functionality will be crucial in the upcoming surveys where each night thousands of transients may be discovered.

  \begin{table}
  \centering
  \caption{Ia vs. non-Ia classification accuracy using baseline RNN trained with whole {\it SALT2 fitted} dataset, a fraction of $0.43$ and the whole {\it complete} dataset. Accuracy for partial light-curve classification with respect of days before or after simulated supernova peak, all indicates all data points available for each light-curve. The addition of host redshift features is indicated as photometric (zpho) or spectroscopic (zspe). For the latter, since not all supernovae have a spectroscopic redshift, we show the accuracy of the subsample with spectroscopic host redshift.}
  \label{table:partial-lcs}
\begin{tabular}{l  cccc }
\multicolumn{5}{c}{SALT2 fitted dataset} \\
\hline
redshift &                -2 &                 0 &                +2 &               all \\
\hline
    None &  $83.6 \pm 0.6$ &  $85.0 \pm 0.7$ &  $86.4 \pm 0.7$ &  $96.3 \pm 0.4$ \\
    zpho &  $93.2 \pm 0.4$ &  $93.8 \pm 0.5$ &  $94.5 \pm 0.5$ &   $98.4 \pm 0.3$ \\
    zspe &  $97.0 \pm 0.2$ &  $97.4 \pm 0.2$ &  $97.9 \pm 0.2$ &  $\mathbf{99.55 \pm 0.06}$ \\
\hline
\end{tabular}
\begin{tabular}{l  cccc }
\multicolumn{5}{c}{$43 \%$ of complete dataset} \\
\hline
redshift &                -2 &                 0 &                +2 &               all \\
\hline
    None &   $86.1 \pm 0.1$ &  $87.24 \pm 0.08$ &  $88.31 \pm 0.09$ &  $96.65 \pm 0.05$ \\
    zpho &   $93.0 \pm 0.3$ &  $93.7 \pm 0.3$ &  $94.3 \pm 0.3$ &  $\mathbf{98.6 \pm 0.2}$ \\
    zspe &  $92.7 \pm 0.4$ &   $93.4 \pm 0.4$ &  $94.0 \pm 0.5$ &   $98.1 \pm 0.2$ \\
\hline
\end{tabular}
\begin{tabular}{l  cccc }
\multicolumn{5}{c}{Complete dataset} \\
\hline
redshift &                -2 &                 0 &                +2 &               all \\
\hline
    None &  $86.4 \pm 0.1$ &  $87.6 \pm 0.1$ &  $88.6 \pm 0.1$ &  $96.92 \pm 0.09$ \\
    zpho &  $93.5 \pm 0.1$ &  $94.2 \pm 0.1$ &  $94.8 \pm 0.1$ &  $\mathbf{98.85 \pm 0.04}$ \\
    zspe &  $93.3 \pm 0.1$ &  $94.0 \pm 0.1$ &  $94.6 \pm 0.1$ &  $98.43 \pm 0.08$ \\
\hline
\end{tabular}
\end{table}
    
\subsection{Classifying many supernovae types}
There is more to supernovae classification than binary classification. Time-domain surveys are increasingly exploring the diversity of supernovae and would benefit from classification of multiple supernova classes. We explore ternary (Ia, Ibc and IIs) and seven-way (Ia, IIP, IIn, IIL1, IIL2, Ib, Ic) classification tasks. We train with the {\it complete} dataset to obtain a large number of light-curves per target.

For ternary classification, we train with $318,820$ light-curves per type and for the seven-way classification with $104,158$ per type. Accuracies for these classifications with and without redshifts are shown in Table~\ref{table:partial-lcs-multiclass}. For complete light-curves our method yields unprecedented classification accuracy, providing a useful tool for obtaining photometric samples of a diversity of supernovae. Early classification becomes a much more challenging tasks and we consequently observe a notable performance degradation. Nonetheless, our algorithm provides a reasonable indication of the possible supernova type and performance is enhanced with the incorporation of redshift information.

\begin{figure}
	\includegraphics[width=1.1\columnwidth]{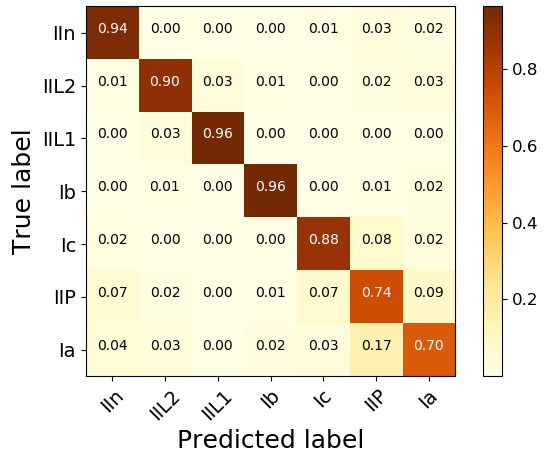}
    \caption{Confusion Matrix for seven-way classification with best performing Baseline classifier with no redshift information. Matrix shows the fraction of light-curves for a given supernova type (True label) classified as the supernova type in the horizontal axis (Predicted label). The diagonal elements are those light-curves correctly classified while off-diagonal elements are those that are mislabeled by the classifier. Color bar indicates the normalized percentage of a certain type of SN light-curves in the predicted label. For seven-way classification our algorithm performs superbly for types Ib, IIL1 and IIn SNe while SNe Ia are not as well characterized.
    }
    \label{figure:cnf_matrix}
\end{figure}

For an equivalent training sample per type, the ternary or seven-way classification accuracy with or without redshift and whole light-curves is much lower than for binary classification (Ia vs. non Ia) as shown in Figure~\ref{fig:accuracy_methods}. Splitting the core-collapse supernovae in subclasses adds a new level of complexity which accounts for the performance drop. Interestingly, for seven-way classification the contamination of the SN Ia sample is dominated by type IIP SN as seen in Figure~\ref{figure:cnf_matrix}. This was seen for early classification in previous Section~\ref{sec:partial-classification} but not for the complete light-curve classification.

As seen in Table~\ref{table:partial-lcs-multiclass}, early classification is severely impacted by adding more classification targets. A thorough study on mechanisms to improve multiclass predictions is out of the scope of this paper but an interesting avenue for further studies. It is possible that a two-step procedure is useful, where: first a multi-target prediction would identify the most probable targets and then, a second prediction with an algorithm specifically trained on the top candidates would refine the classification. The extensive results presented here show that SuperNNova can be a valuable tool for present and future surveys that wish to prioritize spectroscopic and photometric follow-up targets.

    \begin{table}
  \centering
  \caption{Ternary (Ia, Ibc, IIs) and seven-way (Ia, IIP, IIn, IIL1, IIL2, Ib, Ic) classification using baseline RNN trained with complete dataset. Accuracy for partial light-curve classification with respect of days before or after simulated supernova peak. The addition of host redshift features is indicated as photometric (zpho) or spectroscopic (zspe). We evaluate the accuracy for the complete validation sample and for the subset of light-curves that possess a spectroscopic host-galaxy redshift.}
  \label{table:partial-lcs-multiclass}
\begin{tabular}{l  cccc }
\multicolumn{5}{c}{Ternary classification} \\
\hline
redshift &                -2 &                 0 &                +2 &               all \\
\hline
    None &  $69.2 \pm 0.2$ &  $71.6 \pm 0.3$ &  $74.1 \pm 0.3$ &  $92.4 \pm 0.3$ \\
    zpho &  $79.7 \pm 0.1$ &  $81.5 \pm 0.2$ &  $83.4 \pm 0.2$ &  $\mathbf{95.4 \pm 0.1}$ \\
    zspe &  $77.8 \pm 1.1$ &   $79.6 \pm 1.2$ &  $81.5 \pm 1.3$ &  $94.1 \pm 1.0$ \\
\hline
\end{tabular}
\begin{tabular}{l  cccc }
\multicolumn{5}{c}{Seven-way classification} \\
\hline
redshift &                -2 &                 0 &                +2 &               all \\
\hline
    None &  $57.2 \pm 0.3$ &  $60.0 \pm 0.4$ &   $62.9 \pm 0.3$ &  $86.8 \pm 0.3$ \\
    zpho &  $64.5 \pm 0.2$ &  $67.1 \pm 0.2$ &  $69.8 \pm 0.2$ &  $90.0 \pm 0.1$ \\
    zspe &  $64.2 \pm 0.3$ &  $67.0 \pm 0.4$ &  $69.8 \pm 0.4$ &  $\mathbf{90.4 \pm 0.4}$ \\
\hline
\end{tabular}
\end{table}

\section{Bayesian RNNs (BRNNs)}\label{section:BayesianRNNs}
In this Section we introduce Bayesian Recurrent Neural Networks. These RNNs fit a posterior distribution on the weights of the neural network thus allowing us to sample different predictions for a given input. Both of our BRNNs are derived from a technique called \textit{Variational Inference} which we will now quickly review.

\subsection{Variational Inference}

Following \citep{Blundell:2015}, we can view neural networks as a model aiming to correctly estimate $\mathcal{P}\left(\mathbf{y}\vert\mathbf{x}, \mathbf{w}\right)$ where in our case, $\mathcal{P}$ is a categorical distribution, $\mathbf{y}$ is the classification target, $\mathbf{x}$ is the photometric light-curve and $\mathbf{w}$ are the network's weights. Neural networks are traditionally trained using a maximum likelihood criterion: given a set of $N$ labeled training observations, the dataset $\mathcal{D}$ is defined by $\mathcal{D} = \left(\mathbf{x}_i, \mathbf{y}_i\right)_{i=1...N}$, we minimize the negative log likelihood $NLL = \min\limits_{\mathbf{w}} \sum_{i=1}^{N}-\log\mathcal{P}(\mathbf{y}_i \vert \mathbf{x}_i, \mathbf{w})$ with gradient descent.

Moving on to the Bayesian picture, instead of having a fixed value for each weight in the neural network,  we now assign a \textit{distribution} to each weight. We seek to find the \textit{posterior} distribution of the weights: $\mathcal{P}\left(\mathbf{w}\vert \mathcal{D}\right)$ which will then allow us to make predictions for new, new observations $x$: $\mathcal{P}(\mathbf{\hat{y}} \vert \mathbf{x}) = \int \mathcal{P}(\mathbf{\hat{y}}\vert \mathbf{x}, \mathbf{w}) \mathcal{P}\left(\mathbf{w}\vert \mathcal{D}\right)d\mathbf{w}$. Typically, the \textit{posterior} distribution is intractable for deep neural networks. To sidestep this difficulty, we can approximate the posterior with a simple parametric distribution $q(\mathbf{w}\vert \theta)$ called the \textit{variational distribution}. In the gaussian case, $\theta = (\mu, \sigma)$, respectively the mean and standard deviation of this variational distribution. Variational inference introduces a new learning criterion, where one seeks to minimize a divergence (typically, the Kullback-Leibler divergence or KL) between the \textit{posterior distribution} and its \textit{variational approximation}:

\begin{align}
  &\hat{\theta} = \min\limits_{\theta} \mathbf{KL}\left( q(\mathbf{w\vert\theta})\vert\vert \mathcal{P}(\mathbf{w}\vert\mathcal{D}) \right) \\
  &= \min\limits_{\theta} \left[\mathbf{KL}\left( q(\mathbf{w\vert\theta})\vert\vert \mathcal{P}(\mathbf{w}) \right) - \int
  q(\mathbf{w}\vert\theta)\log\mathcal{P}(\mathcal{D}\vert\mathbf{w}) d\mathbf{w}  \right] \\
  &= \min\limits_{\theta} \left[\mathbf{KL}\left( q(\mathbf{w\vert\theta})\vert\vert \mathcal{P}(\mathbf{w}) \right) - \mathbb{E}_{q(\mathbf{w}\vert\theta)}(\log\mathcal{P}(\mathcal{D}\vert\mathbf{w})) \right]
  \label{eq:B_loss}
\end{align}

where $\mathcal{P}(\mathbf{w})$ is a user-specified \textit{prior distribution} over the neural network's weights.
This new cost function is made up of two terms. The first one, the KL term, is a regularization term: it penalizes variational distributions which differ too much from the prior. The second one is a likelihood term; our model must be flexible enough to handle the complexity of the data distribution. Bayesian optimization of neural networks is a trade-off between those two terms.

We will now review two ways in which the variational distribution can be specified and investigate their applications to supernova cosmology. Results in the following sections use the {\it complete dataset} for training, validation and testing.

\subsection{MC dropout}\label{section:variational}

Following \citep{Gal_a:2015,Gal_b:2015} we define our variational distribution to factorize over each row $\mathcal{r}$ of the network's weight matrices: $q(\mathbf{w}_r) = p \mathcal{N}(0, \sigma^2I) + (1 - d)\mathcal{N}(\mathbf{\mu}_r, \sigma^2I)$ where $\mathbf{w}_r$  is the weight matrices rows, $d$ is the dropout probability, $\mathcal{N}$ the normal distribution, $\mathbf{\mu}_r$ is the variational parameter (row vector) optimised with our gradient descent algorithm, $\sigma^2$ is a fixed, small constant and $I$ the identity. Using a normal prior on the weights, the KL term can be approximated as $L_2$ regularization on the network's weights. Evaluating this network then becomes equivalent to performing dropout (i.e. masking with zeros) on the rows of the weight matrices (or equivalently, on each layer's input). The network can then be trained as usual, as long as we use the same dropout mask at every time step in the sequence \citep{Gal_a:2015,Gal_b:2015}.

We can now obtain a distribution of predictions, simply by sampling a different dropout mask for each prediction. In this work we sample predictions fifty times for each light-curve. The median of the prediction array is used to report the accuracy score. These uncertainties can provide valuable information in the classifier's confidence on the prediction.

 We use the same hyper-parameters as those of the baseline RNN presented in Section~\ref{section:BaselineRNN}. We probe different dropout values, $p \in [0.01,0.05,0.1,0.2]$ and weight decay of the gradient descent optimiser $\in [1e^{-5}, 1e^{-7},1e^{-9}]$ to evaluate the performance of the network with a data fraction of $0.2$. We find that the two most accurate architectures have a dropout of $0.01$ and a weight decay of $1e^{-5}$ and $1e^{-7}$. Since accuracies for both models differ by less than $0.1\%$, we choose to use weight decay $1e^{-7}$ in the following. For the {\it complete} dataset this configuration has an accuracy of $96.87 \pm  0.09$ without redshift and $98.8 \pm 0.1$ with host photometric redshift for complete light-curve classification which is equivalent to the baseline RNN accuracy with a slight reduction. We find that the efficiency ($98.91 \pm 0.05$), purity ($98.6 \pm  0.2$) and contamination values are similar but slightly less than to those of the Baseline implementation. In the next Section~\ref{section:towardscosmology}, we will discuss other criteria for selecting the best performing method, through calibration and entropy of Out-of-Distribution events.

\subsection{Bayes by Backprop (BBB)}\label{section:BBB}

Following \cite{Fortunato:2017}, to model the variational distribution $q(\mathbf{w}\vert \theta)$ we use a simple Gaussian distribution with mean $\mu$ and standard deviation $\sigma$.

During training, we sample each weight $\mathbf{w}$ as follows: sample $\boldsymbol{\upvarepsilon} \sim \mathcal{N}(0, 1)$ then obtain $\mathbf{w} = \boldsymbol{\mu} + \boldsymbol{\sigma} \times \boldsymbol{\upvarepsilon}$. Where $\boldsymbol{\sigma},\boldsymbol{\mu}$ are optimized by gradient descent. 

Following \cite{Blundell:2015}, we partition the training data in $n_B$ minibatches $\mathcal{D} = (\mathcal{D}_j)_{j=1...{n_B}}$. For each batch, we optimize:
\begin{align}
\mathcal{L}_j = \frac{1}{n_B}  \mathbf{KL}\left( q(\mathbf{w\vert\theta})\vert\vert \mathcal{P}(\mathbf{w}) \right) - \mathbb{E}_{q(\mathbf{w}\vert\theta)}(\log\mathcal{P}(\mathcal{D}_j\vert\mathbf{w})) 
\end{align}
where the KL-divergence is estimated with Monte Carlo sampling and the prior is given by a mixture of Gaussians with standard deviation $\sigma_1, \sigma_2$ respectively as: $\mathcal{P}(\mathbf{w}) = \pi \mathcal{N}(0, \sigma_1) + (1 - \pi) \mathcal{N}(0, \sigma_2)$. $ \sum_{j=1}^{n_B} \mathcal{L}_j$ is equivalent to the loss in Eq.~\ref{eq:B_loss}.\\

Using the same hyper-parameters as those of the baseline RNN presented in Section~\ref{section:BaselineRNN}, we study the impact of modifying the prior with $0.2$ of the {\it SALT2 fitted} dataset. We set $\pi = 0.75$. We searched $\log\sigma_{1}\in\left\{-2, -1 \right\}$, $\log\sigma_{2} \in\left\{-4, -7\right\}$ for the recurrent layer and $\log\sigma_{1}\in\left\{-2, -1, -0.5 \right\}$, $\log\sigma_{2} \in\left\{-1, -0.5, -0.1 \right\}$ for the output layer. We initialized the standard deviation of the weights of the recurrent layer by sampling uniformly between $\log \left( e^{\frac{\sigma_{\rm mix}}{\rho_{\rm lower}}} - 1\right)$ and $\log \left( e^{\frac{\sigma_{\rm mix}}{\rho_{\rm upper}}} - 1\right)$, where $\sigma_{\rm mix} =\sqrt{\pi \sigma_1^2 + (1 - \pi)\sigma_2^2}$, $\rho_{\rm lower} \in \left\{4, 3\right\}$ and $\rho_{\rm lower} \in \left\{2, 1\right\}$ for the recurrent layer and $\rho_{\rm lower} \in \left\{3,2,1\right\}$ and $\rho_{\rm lower} \in \left\{2, 1\right\}$ for the output layer. 

We find that the Bayesian network trains and attains better performance without the cyclic learning rate. 

The most accurate model has a $\log\sigma_1=-1$, $\log\sigma_2 = -7$, $\rho_{\rm lower} = 4$ and $\rho_{\rm upper} = 3$ for the recurrent layer and $\log\sigma_1=-0.5$, $\log\sigma_2 = -0.1$, $\rho_{\rm lower} = 3.0$ and $\rho_{\rm upper} = 2.0$ for the output layer. This model reaches an accuracy of $ 96.85 \pm  0.05$ without redshift and $98.72 \pm 0.04$ with host photometric redshift information for whole light-curve classification in the {\it complete} dataset. The efficiency, purity and contamination values are similar to those of the MC dropout implementation and slightly reduced with respect to the Baseline implementation $\leq 0.5\%$. In Section~\ref{section:towardscosmology}, we will discuss other criteria for selecting the best performing method, through calibration, entropy of Out-of-Distribution events and model uncertainties. 

\subsection{BRNNs Uncertainties}\label{section:BRNN_uncertainties}
In classification tasks, evaluating the reliability of a model's uncertainties is of paramount importance. Following \cite{Kendall:2017}, we split uncertainties into two categories: {\it aleatoric} and  {\it epistemic}. {\it Aleatoric} uncertainties capture the uncertainties in the input data, e.g. noise or other effects of data acquisition. {\it Epistemic} uncertainties express our ignorance about the model that generated the data. The latter depends on the network structure and the training set and therefore can be reduced with more flexible models, as well as larger and more diverse training sets.

In SuperNNova, we do not model {\it aleatoric} uncertainties since those are already included in the dataset (through simulated flux measurement errors). By casting learning in a Bayesian framework, we expect to model {\it epistemic} uncertainties. 

For BRNNs, given a light-curve ${\mathbf{x}}$ we can obtain a set of predictions $p_j(\mathbf{x})_{j=1...n_s}$ where the weights are sampled $n_s$ times from the prediction's posterior. We compute the model uncertainty for a given light-curve $\mathbf{x}_i$ as:
\begin{equation}\label{eq:uncertainty}
\widehat \sigma = std \left( \sum_{j=1}^{n_s} p_{j}({\mathbf{x}_i})\right)
\end{equation}
where $j \in [1,n_s]$ is the index of inference samples, $p_{j}({\mathbf{x}_i})$ is the classification probability for the given light-curve $\mathbf{x}_i$ for each inference sample $j$, and $std$ is the standard deviation. 

\subsubsection{Uncertainty evaluation metrics}\label{section:uncertainties_metrics}
To evaluate the uncertainty estimates of the BRNNs in SuperNNova, we make use of two metrics:

\begin{itemize}
\item{ {\it Mean model uncertainty}: for a given test set $\mathcal{D}_k$, we average the model uncertainties ($ \widehat \sigma $ from Eq.~\ref{eq:uncertainty}) over all light-curves:
\begin{equation}\label{eq:mean_uncertainties}
\langle  \widehat \sigma_{k}  \rangle = \frac{1}{N} \sum_{i=1}^N \widehat \sigma_i .
\end{equation}
where $i\in [1,N]$ is the index of each light-curve in test set $\mathcal{D}_k$.

For two given sets of predictions, we can compute the difference between their uncertainties as:
\begin{equation}\label{eq:delta_sigma}
\Delta \langle \widehat \sigma_{1,2} \rangle = \langle \widehat \sigma_{1} \rangle - \langle \widehat \sigma_{2} \rangle.
\end{equation}
}
where $\widehat \sigma_{i}$ is defined in Equation \ref{eq:mean_uncertainties} for a set of predictions $i$.

\item{  {\it Entropy}: we follow \cite{Fortunato:2017} and define, for a test dataset $ \mathcal{D}_{t}:[\mathbf{x}_1,...\mathbf{x}_N]$ with $N$ light-curves and a classification model $m$, the entropy of $ \mathcal{D}_{t}$ under $m$ as:
\begin{equation}
H_m[\mathcal{D}_{t}] = \sum_{i=1}^N p_m(\mathbf{y}_i|\mathbf{x}_i) log \left( \frac{1}{p_m (\mathbf{y}_i|\mathbf{x}_i)} \right).
\end{equation} \label{eq:entropy}
where $p_m(\mathbf{y}_i|\mathbf{x}_i)$ is the classification probability given the light-curve $i$ using model $m$. Entropy is a proxy for the model's confidence on predictions. Thus, confident predictions will yield low entropy. An equivalent per light-curve entropy can be computed as $\overline{H_m}[\mathcal{D}_{t}] = \frac{1}{N} H_m[\mathcal{D}_{t}]$, where $N$ is the number of light-curves in a given test set. For two given set of predictions, we can define the {\it entropy gap} $\Delta H$ by:
\begin{equation}\label{eq:delta_entropy}
\Delta H = \quad \overline{H_{m_1}}[\mathcal{D}_{t}] - \overline{H_{m_2}}[\mathcal{D}_{t}]
\quad or = \overline{H_m}[\mathcal{D}_{t_1}] - \overline{H_m}[\mathcal{D}_{t_2}].
\end{equation}
where the first option evaluates the entropy gap over the same dataset for two given models ($m_1,m_2$), and the second uses the same model to make predictions for two different data sets ($ \mathcal{D}_{t_1},\mathcal{D}_{t_2}$). }
\end{itemize}

\subsubsection{Uncertainty evaluation}

We now evaluate the uncertainty estimates of the BRNNs in SuperNNova. Bayesian Neural Networks aim to capture {\it epistemic uncertainties} by putting a prior distribution over the NN's weights. Bayesian inference leads to computing a posterior which represents the set of plausible models, given the data. As more data is available, we expect these uncertainties to be explained away. To verify this, we make predictions with two models on the same test set. The models differ only by the number of light-curves used for training. 

First, we compute $\Delta \langle \widehat \sigma \rangle$ (Eq.~\ref{eq:delta_sigma}) between the predictions obtained with a model trained with a small number of light-curves and one trained with a larger number. We expect this metric to be positive. For the {\it complete} dataset we compute the $\Delta \langle \widehat \sigma \rangle$ of models trained with $43\%$ and the whole training set, $\Delta \langle \widehat \sigma \rangle =  \langle \widehat \sigma_{0.43}\rangle - \langle \widehat \sigma_{whole} \rangle $. We find $\Delta \langle \widehat \sigma \rangle = 0.005$ and $0.007$ respectively. We find similar values for the $\Delta \langle \widehat \sigma \rangle$ of models trained with half and the {\it salt fitted} training set.

Second, we compute the entropy gap defined in Eq.~\ref{eq:delta_entropy} between the same predictions. Since we expect the predictions from the model trained with less light-curves to be more uncertain than the one from a model trained with a larger dataset, we should obtain a positive $\Delta H$. We find for the MC dropout and BBB implementations a positive $\Delta H > 0.01$.

Both BRNNs implementations have uncertainties which are consistent with the behavior expected of {\it epistemic} uncertainties as shown by the metrics computed above. However, we find that the size of uncertainties differ in both methods. If we compute the mean of the classification uncertainties when classifying the {\it complete} dataset, we find that the MC dropout implementation has uncertainties twice larger than the BBB implementation. This may be due to initialization effects or more fundamental effects due to the way each method specifies variational distributions. Future research should strive to improve the comprehension of these uncertainty estimates.

In the following Section, we will further study the behavior of our BRNNs uncertainties. In particular, we will explore the effect of non-representative training sets and the classification of out-of-distribution light-curves.
     
 \begin{figure}
\includegraphics[width=1.05\columnwidth]{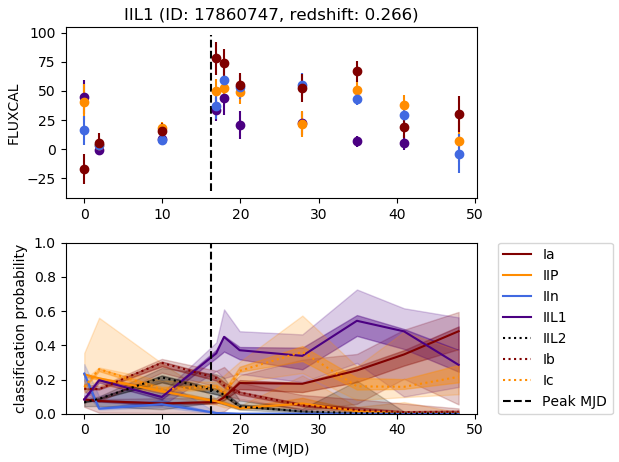}
\caption{Classification of simulated type IIL1 SN as type Ia SN with seven-way MC dropout RNN. The model was trained without redshift information. Predictions for this light-curve have low confidence and have large uncertainties (shadowed regions show $68\%$ and $95\%$ contours). The large uncertainties on the classification probability are representing the lack of confidence in this classification and are correlated to the low S/N.}
\label{figure:variational_multiclass_missclassified}
\end{figure}
   
 \section{Towards cosmology and other statistical analyses}\label{section:towardscosmology}
 To perform statistical analyses using photometrically classified supernovae, a high accuracy algorithm is not enough. It is equally important to show that it is statistically sound. By that, we mean that it should provide well calibrated probabilities and capture the {\it epistemic} uncertainties related to the classification model. In the following, we will quantify the performance of our algorithms with respect to these requirements focusing on the Ia vs. non Ia classification task.

 \subsection{Calibration} \label{section:calibration}
Classification probabilities should reflect the real likelihood of events being correctly assigned to a target. Classification algorithms where this is true are said to be calibrated. \cite{Niculescu:2005} show that common machine learning algorithms such as SVMs and boosted trees do not predict well calibrated probabilities, pushing predicted probabilities away from $0$ and $1$. For other algorithms such as Random Forest, the calibration is heavily data dependent. Recently, \cite{Guo:2017} showed that modern, deep neural networks also suffer from poor calibration and that there is a trade-off between classification performance and calibration.

To analyze our algorithms' calibration, we use reliability diagrams \citep{DeGroot:1983}. These diagrams are constructed by discretizing the predicted probability into ten evenly spaced bins. A predicted probability between $0.0$ and $0.1$ falls into the first bin, and so on. For each bin, we plot the fraction of true positive cases against the mean predicted probability in that bin. The fraction of true positive cases in the binary classification is defined by the number of type Ia supernovae in that probability bin with respect to all supernovae in that bin. If the model is well calibrated, the points will fall near the diagonal line. This is equivalent to saying that in a sample with a hundred events classified as type Ia with probability $0.7$, we expect $70\%$ of events to be true SNe Ia and $30\%$ to be missclassified SNe from other types. Furthermore, we construct a metric to study the calibration deviation: the difference between the two calibrations squared. 

\begin{figure}
\includegraphics[width=1\columnwidth]{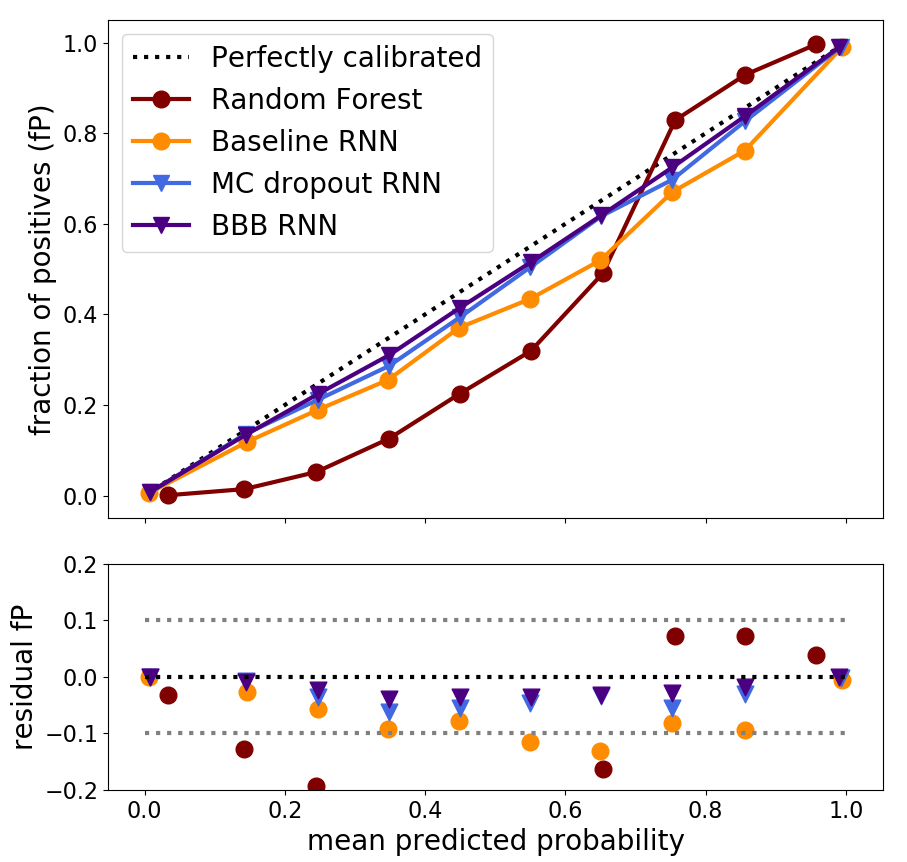}
\caption{Calibration of classification algorithms. Top: reliability diagram showing the calibration for {\it SALT2 fitted} dataset classification for a single seed. We use the most accurate configurations for the Random Forest (red circles), Baseline RNN (yellow circles), MC dropout RNN (blue triangles) and BBB RNNs (purple triangles). Bottom: dispersion from perfectly calibrated algorithms. Note that the Random Forest algorithm has a large deviation from perfect calibration while the RNNs are better calibrated than this algorithm with the BBB implementation almost perfectly calibrated.}
\label{figure:cal_salt}
\end{figure}

For the Random Forest algorithm, we find a large calibration deviation when classifying the {\it SALT2 fitted dataset}, as can be seen in Figure~\ref{figure:cal_salt}. Over five randomized runs the level of dispersion is found to be $0.025 \pm 0.002$. In this classification task, RNNs are found to have better calibration than the Random Forest algorithm with a dispersion an order of magnitude lower. For BBB and MC dropout RNNs we construct reliability diagrams using multiple predictions per sample, rather than the median prediction. We find that diagrams built this way exhibit better calibration than those built with a single prediction per sample. This is evidence that the model has learned meaningful predictive uncertainties since including the complete distribution of probabilities improves the calibration. 

Bayesian RNNs are found to be better calibrated than the Baseline RNN for both {\it salt fitted} and {\it complete} datasets. For the classification of the {\it complete} dataset without any redshift information, we find a calibration dispersion from the Baseline RNN of $0.006 \pm 0.001$ which is reduced to $0.004 \pm 0.002 $ for the MC dropout and to $0.0005 \pm  0.0004$ for the BBB implementations.  

Calibration depends on the nature and size of the training set. We verify this, by measuring the dispersion for the Baseline RNN when classifying the {\it SALT2 fitted dataset} without redshift information with data fractions between $\{0.2-1.0\}$. For the nature of the training set, we compare using the whole {\it SALT2 fitted} dataset and $0.43$ of the {\it complete} dataset. We find that the data fraction or nature of the dataset can change the dispersion up to $50\%$.

Photometrically classified samples are usually selected from those events that have a probability larger than a given threshold. These thresholds are chosen as a compromise between purity and size of the selected sample. However, miscalibration affects the positive fraction of events in each bin, providing misleading probabilities. To account for large calibration deviation, two approaches may be taken: either to perform a post-processing recalibration (e.g. \cite{Niculescu:2005,Guo:2017}) or the difference between the obtained and true probability for each bin can be used to re-weight obtained probabilities. This will be of importance for classifier and datasets where large calibration dispersion is observed. We consider our BBB and MC dropout RNNs to be well calibrated due to deviations less than one percent.

 \subsection{Representativeness}\label{section:representative}
To improve generalization, classification algorithms typically require large and representative training sets. A training set is called representative if its properties, such as maximum observed brightness, have distributions that resemble the distributions of the dataset to classify also known as a test set. Training sets are usually simulated using templates from spectroscopically followed up supernovae. Current and past surveys have spectroscopic follow-up strategies designed to target as many Ia-like objects as possible. These spectroscopically selected samples are therefore not probing the complete supernovae diversity. In this section, we investigate the impact of training SuperNNova with non-representative and representative datasets.

Previous works have discussed the issue of representativeness by comparing the accuracy of an algorithm trained with a dataset whose size was similar to the test set \citep{Lochner:2016,Charnock:2017,Pasquet:2019}. RNN based methods, trained on small datasets, are prone to over-fitting \citep{Charnock:2017}. Additionally, we have shown that the performance of our algorithm depends on the size of the training set (see Figure~\ref{fig:accuracy_methods}). Thus, it would not be adequate to use this procedure to evaluate the impact of representativeness.

\begin{figure}
\includegraphics[width=1\columnwidth]{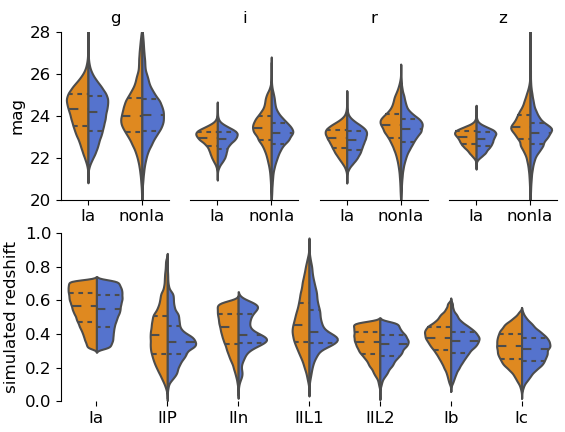}
\caption{Distributions of maximum observed brightness (mag), in all DES filters ($g,i,r,z$), and of simulated redshift for {\it SALT2 fitted} (left yellow) and {\it complete} (right blue) datasets. Maximum observed brightness is shown for type Ia and non-Ia samples while simulated redshift is shown for each of the available templates. Dashed lines show the median and first quartile of the distribution. The {\it complete} and {\it SALT2 fitted} datasets probe similar parameter space but have different distributions. This is similar to what is expected of non-representative samples. }
\label{fig:violin}
\end{figure}

Instead, we use the following approach. We select the {\it SALT2 fitted} dataset as an example of non-representative data. Recall that this dataset contains light-curves successfully fitted by SALT2 and is therefore biased towards light-curves that look like SNe Ia. As representative data we select the {\it complete} dataset which includes all simulated light-curves "discovered" by the DES-like survey and pass selection cuts. This sample has larger diversity of light-curves and is thus a better probe of the full data distribution space. The distribution of magnitudes and redshifts for both datasets is shown in Figure~\ref{fig:violin}. To investigate the discrepancy between spectroscopic and photometric samples, we train with the non-representative dataset and evaluate the classification performance for the test sample in the representative dataset.

For classification with no redshifts using the Baseline and Bayesian RNNs we find that the accuracy is reduced by $1\%$ when trained with a non-representative dataset. Although small, this variation is not within the uncertainties of our model accuracies.

As discussed in Section~\ref{section:BRNN_uncertainties}, Bayesian RNNs can capture {\it epistemic uncertainty} which includes the lack of diversity in the model's training set. Therefore, we expect a Bayesian RNN trained on non-representative set (in our case, the full {\it SALT2 fitted} dataset) to be more uncertain than one trained on a representative set (in our case, a subset of the {\it complete} dataset) when evaluating on said representative set. To quantify this in a rigorous way, we use the two metrics introduced in Section~\ref{section:uncertainties_metrics}. We find both metrics to be positive for all BRNNs, $\Delta \langle \widehat \sigma \rangle  > 0.004$ and $\Delta H >  0.01$. 

The lack of representativity and the limitations of supernova templates are major issues in SN photometric classification. Recently, \cite{Ishida:2018} introduced a framework for the optimization of spectroscopic follow-up resources to improve SN photometric classification datasets. Bayesian RNN uncertainties may be a promising indicator to select follow-up candidates for this purpose. This is an interesting possibility which we leave for future work.

\subsection{Out of distribution light-curves (OOD)} \label{section:ood}
In astronomy, as in any other classification application, the generalization properties of a classifier and its behavior on unseen, possibly out-of-distribution samples represents a challenge. In this section we study the performance of SuperNNova when classifying out-of-distribution light-curves. We test four different types of OODs: time reversed light-curves, randomly shuffled light-curves, random fluxes and a sinusoidal signal. The latter two were generated using the same cadence and flux range as normal supernovae. These light-curves are only used for testing and were not used for training at any time.

\begin{figure}
\includegraphics[width=1.05\columnwidth]{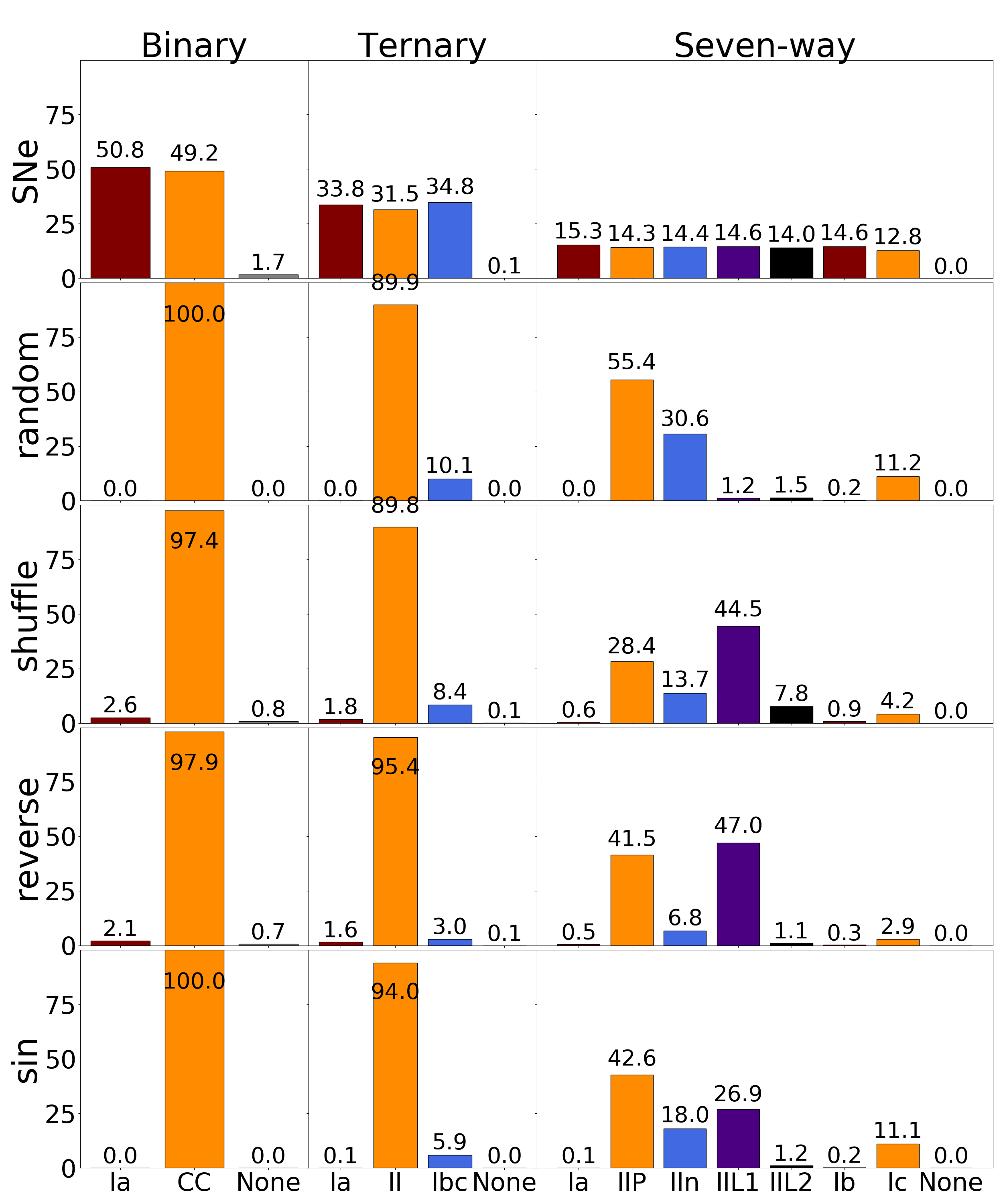}
\caption{Predicted classification percentages for best configuration BBB RNN with single seed. Columns indicate number of targets for the classification (with probabilities summing one) and an additional column indicating a low probability prediction, $<0.6,0.4,0.2$ for 2, 3 and 7 classes respectively. Rows represent the different light-curve types. All rows are out-of-distribution generated light-curves except the row ``SNe" representing our testing dataset. Since the testing set has balanced classes, we expect the prediction percentage in the row ``SNe" to be balanced as well. Note that the out-of-distribution events are rarely classified as type Ia SNe ($<2.6\%$ binary, $<1.8\%$ ternary and $<0.6\%$ in seven-way classification), the highest percentages in binary classification are for OOD which can resemble SNe light-curves, e.g. reverse and shuffle. In ternary and seven-way classification, out-of-distribution events are preferably classified as type II, IIn or IIL1 SNe. Note that the number of low probability detections are much higher for OOD when classifying in three or seven classes.}
\label{figure:ood_bar}
\end{figure}

When classifying out-of-distribution light-curves, all SuperNNova algorithms rarely classify these light-curves as SNe Ia. For binary classification, for the MC dropout and BBB implementation, the reverse and shuffle light-curves obtain the largest number of classifications as SNe Ia, $4.8\%$ and $4.1\%$ respectively with the MC dropout implementation and less than $3\%$ for the BBB. Many of these light-curves resemble supernovae, specially with light-curves with low signal-to-noise. For the sinusoidal light-curves, only the MC dropout classifier classifies a significant percentage as SNe Ia $5.9\%$, while for the BBB method this is not the case. The network seems to characterize type Ia supernovae well and therefore classifies most OOD events as core-collapse supernovae as can be seen in Figure~\ref{figure:ood_bar} with the BBB RNN. In ternary and seven-way classification the most common predictions for OOD events are types II: IIn, IIp, IIL1. 

\begin{figure}
	\includegraphics[width=1.05\columnwidth]{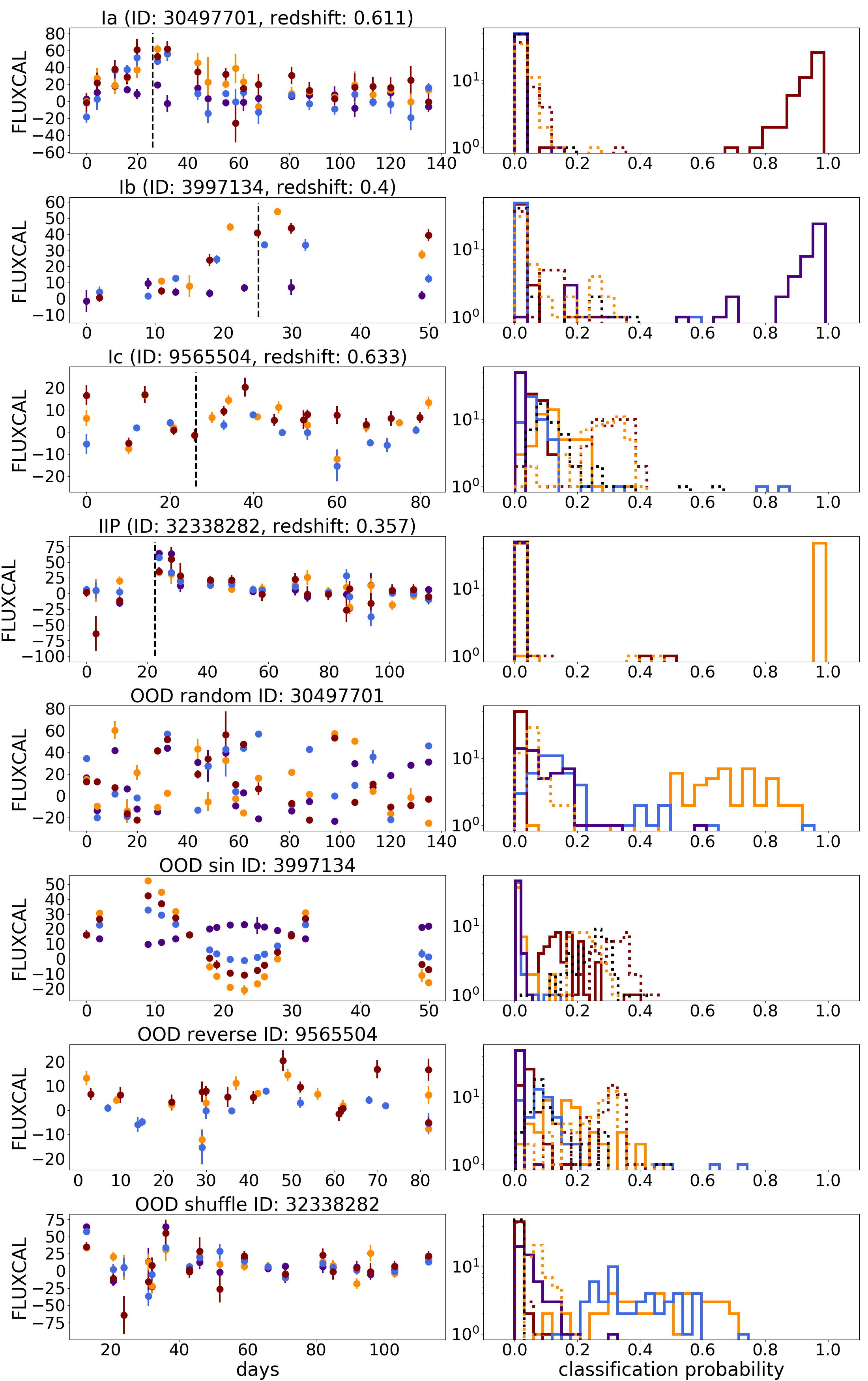}
    \caption{Out-of-distribution and SN light-curve prediction distribution for best MC dropout configuration without redshift information for seven-way classification. First column depict light-curves, second column their classification probabilities for different supernova classes depicted with various line styles and colors. First four rows show SNe light-curves while last four rows are out-of-distribution events. For the OOD events we see different behaviours: high classification probabilities for a certain supernova class with large uncertainties, clustered classification probabilities around a small value and a mixture of both behaviours.}
    \label{figure:prediction_distribution}
\end{figure}

To assess how our prediction uncertainties behave with respect to out-of-distribution events, we compute $\Delta H$ for our best performing models comparing OOD and our {\it complete} data test set predictions. Although $\Delta H$ does not measure uncertainty in an absolute way, it can be used as a qualitative test where OOD events should show high entropy as seen in \cite{Fortunato:2017}. For the binary classification problem, we find positive $\Delta H$ for random and reverse light-curves for both the MC dropout and BBB implementations, with the BBB having the largest entropy gap. Interestingly, for sinusoidal and shuffle light-curves this metric is negative for the MC dropout implementation. For the ternary and seven-way problems, other OOD predictions are as well negative.While \cite{Fortunato:2017} observe large positive entropy for OOD events, our experiments show surprisingly a mixed behaviour. We explore this question in Appendix~\ref{appendix:mnist} where we conclude that RNNs are at risk of collapse on predicting a single class with high probability. Additionally, we show the different behaviours for classification of OOD events in a seven-way classification in Figure~\ref{figure:prediction_distribution}.

\subsection{Type Ia supernova cosmology}
Current and future surveys such as DES, PanSTARRS and LSST will use photometrically classified type Ia supernovae for their cosmological analyses. In this section we examine the properties of type Ia supernova samples classified by SuperNNova.

To study of the expansion of the Universe, type Ia supernovae are used to measure distance modulus as a function of redshift. We can compute the distance modulus, $\mu$, of a given type Ia supernova:
\begin{equation} \label{equation:mu}
\mu = m_B +\alpha x_1 - \beta c + M,
\end{equation}
where $M$ is the absolute magnitude of the SNIa, $m_B$ is the rest-frame B magnitude (at peak luminosity), $x_1$ is the stretch parameter and $c$ is the color. The last three parameters are derived from the SALT2 fit to the observed SN Ia light curve. $M$, $\alpha$ and $\beta$ and  can be constrained during cosmological fits. In this work, we will use the simulated values: $M = 19.365$, $\alpha = 0.144$ and $\beta = 3.1$.

With SuperNNova, we obtain a photometrically classified type Ia supernovae sample by selecting light-curves that have a probability larger than $0.5$ to be type Ia. Importantly, we have shown in Section~\ref{section:ood} that out-of-distribution light-curves are rarely classified as type Ia SNe for any of our algorithms. We do not correct calibration deviations found in Section~\ref{section:calibration} given the small miscalibration found which will have a minor effect on the photometrically selected SN Ia sample population which is assessed in this Section. 

\begin{figure}
	\includegraphics[width=\columnwidth]{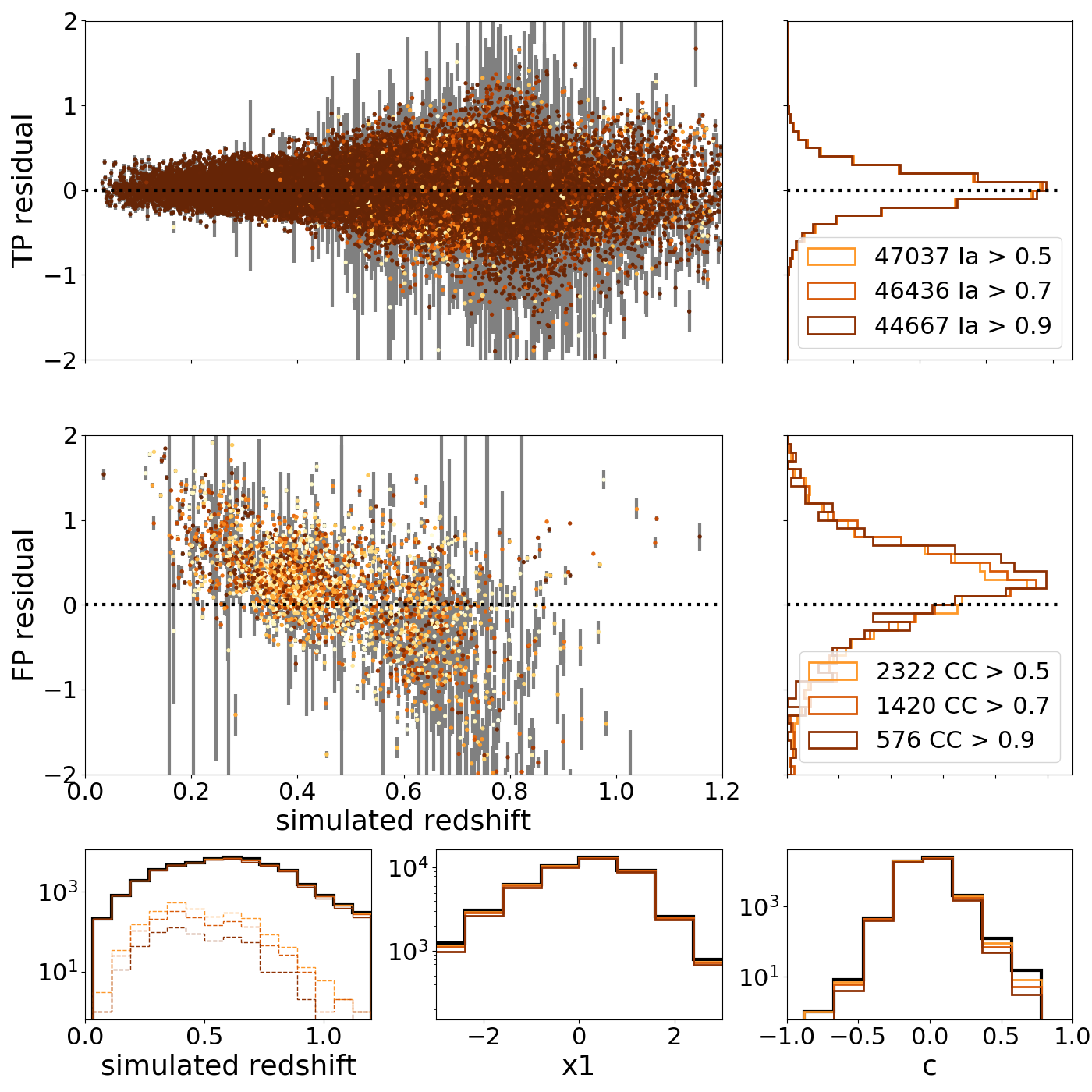}
    \caption{Hubble Residuals as a function of simulated redshift for the photometrically classified SN Ia sample using our best performing BBB RNN trained with the {\it complete} dataset with no redshift information. Top left: Hubble Residuals for correctly classified type Ia supernovae (TP) as a function of simulate redshift. Colors correspond to classification probabilities smoothly distributed between 0.5 (yellow) and 0.9 (dark orange). Top right: normalized histogram of Hubble Residual for three selection thresholds: 0.5 (yellow), 0.7 (orange) and 0.9 (dark orange). Center left and right: same plots as above but for incorrectly classified core-collapse which represent the sample contamination. Bottom from left to right: redshift, color (c) and stretch (x1) distributions of the complete simulated Ia sample (black), selected TP at different probability thresholds (colored continuous lines) and selected FP (dashed lines). We plot only the supernovae classified as type Ia that have had a successful SALT2 fit which provides parameters necessary to compute the distance modulus. Note the small difference between the simulated and selected redshift, color and stretch distributions and the contamination contribution $<5\%$. Clearly, such photometric sample significantly reduces selection biases usually seen in spectroscopically selected samples.}
    \label{figure:HD_BRNN}
\end{figure}

For each supernova in the photometrically selected sample, we obtain its distance modulus using Equation~\ref{equation:mu} and compare it with the $\Lambda CDM$ distance modulus corresponding to that simulated redshift. This is called the Hubble Residual and ideally should be around zero. We show the Hubble Residual for a sample classified with the best-performing BBB RNN without redshift information in Figure~\ref{figure:HD_BRNN} for both correctly classified type Ia supernovae and incorrectly classified core-collapse supernovae. The contribution by the latter represents less than $5\%$ of the sample and peaks at redshifts between $0.4-0.5$. 

Such contamination can be reduced by: using additional information such as host-galaxy redshifts, raising the probability threshold for sample selection as shown in Figure~\ref{figure:HD_BRNN} or eliminating classified events with large uncertainties. We have shown that adding redshifts allows us to obtain a sample contaminated by $<2\%$ core-collapse SNe, contamination which is within the range of current spectroscopically classified samples \citep{Rubin:2015}. Raising the probability threshold is found to be a good alternative, e.g. for the MC dropout implementation, setting the threshold to $>0.9$ increases the sample purity by reducing the number of miss-classified core-collapse SNe by a factor of four and only reducing the correctly classified SN Ia by $<5\%$.

Uncertainties from Bayesian RNN's may provide useful information to discard supernovae in cosmology analyses. They may be incorporated to the analysis of photometric samples with Bayesian Hierarchical Cosmology models such as Steve \citep{Hinton:2018}. As a first toy approach, we discard those SNe classified with uncertainties larger than $0.05$ in the Hubble Diagram, finding a similar decrease in the number of core-collapse SNe as raising the threshold to $0.9$. The number of TP SNe Ia are also equivalent. This shows that uncertainties are a good indicator for possible contamination. A more robust study on the impact of selection by classification probabilities and uncertainties is left for future work.

\begin{figure}
	\includegraphics[width=\columnwidth]{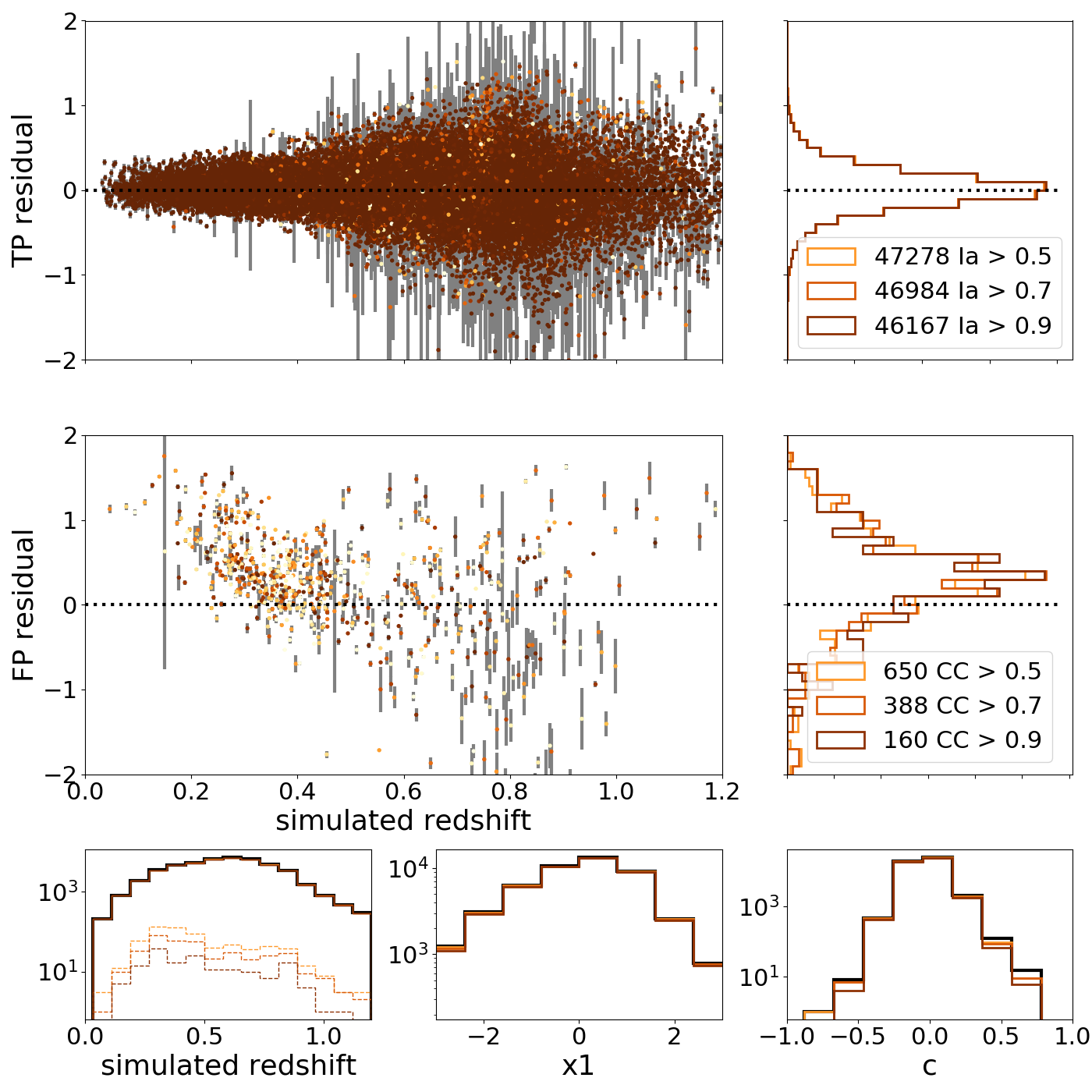}
    \caption{Hubble Residuals as a function of simulated redshift for the photometrically classified SN Ia sample using our best performing BBB RNN with photometric host-galaxy redshift information. Figure is similar to~\ref{figure:HD_BRNN}. Note that the level of contamination obtained in this sample is less than $2\%$ which is equivalent to that expected of spectroscopically classified samples. As in the previous Figure, our selected sample probes almost completely the whole redshift, color and stretch parent population.}
    \label{figure:HD_BRNN_zpho}
\end{figure}

Finally, we examine the population properties of our SuperNNova selected sample when compared to its parent population as can be seen in Figures~\ref{figure:HD_BRNN} and~\ref{figure:HD_BRNN_zpho}. In spectroscopically selected samples, there is a large difference between its properties and the ones of the expected supernovae population. This is mainly due to spectroscopic selection effects \citep{Kessler:2018,Dandrea:2018}. In our photometrically classified sample using no redshift information, we see almost constant selection efficiency throughout redshift. Importantly, color and stretch distributions match very closely the parent population. Asides from this selection bias reduction, we are able to classify type Ia supernovae at redshifts $>0.9$ which is extremely difficult when using spectroscopy. 
    
\section{SuperNNova resources}\label{section:resources}
SuperNNova has been optimized for a reasonable computing budget and special care has been dedicated to memory consumption. For this project we used an 8-core Intel(R) Core(TM) i7-7700K CPU (4.20GHz) with $15$GB of RAM and an NVIDIA GeForce 1080Ti GPU to speed up training and inference.

Database construction, which includes formatting and normalization, is carried out once for the whole dataset and requires less than half an hour. For a given model, training can be carried out on the CPU or, if available, on the GPU. We quote speeds for a batch size of $128$, larger batch sizes will result in a decrease of training time. Similarly, we quote times for training with cyclic learning rate, without cyclic training times increase. For the Baseline RNN we find that training a single epoch with approximately $5,000$ light-curves ranges from $1$ minute (GPU) to $10$ minutes (CPU). Training the Baseline RNN to convergence requires from one and a half hours (GPU) to $15$ hours (CPU). For the same light-curve numbers per epoch, MC dropout RNNs require slightly longer to train with $1.5$ minutes (GPU), while BBB RNNs require $2$ minutes per epoch (GPU).

Once a model is trained, it can be applied to classify independent light-curve samples. SuperNNova is able to classify tenths to hundreds of light-curves per second with BRNNs as can be seen in Table~\ref{table:classification_speeds}. This includes sampling $50$ times the probability distribution ($50$ being the number of inference samples). For the Baseline RNN (using either GPU and CPU) we are able to classify thousands of light-curves per second. Inference speed will be critical given the unprecedented amounts of data expected from upcoming transient surveys. Thus, SuperNNova paves the way towards real-time light-curve classification.
    
\begin{table}
\centering
\caption{Classification speeds for batch sizes of 128 light-curves. For each RNN model we measure how many supernovae per second can be classified using either CPU or GPU.}    
\label{table:classification_speeds}
\begin{tabular}{l  cc } 
RNN model & Number of SNe per second & device \\
\hline
Baseline  & 23,280  & GPU \\
Baseline  & 2,820  & CPU \\
MC dropout  & 342  & GPU \\
MC dropout  & 58  & CPU \\
BBB  & 264  & GPU \\
BBB  & 56  & CPU \\
\end{tabular}
\end{table}

\section{Discussion and Conclusions}
We have presented SuperNNova, a fast and accurate framework for photometric classification. SuperNNova only requires light-curves as input. Additional information, such as host-galaxy redshifts, can be easily incorporated to improve its performance. We have released the source code on github alongside with extensive documentation, tests sets and a docker environment to foster exploration and reproducibility \footnote{\url{https://github.com/supernnova/SuperNNova}}.

Current and future surveys will continue to discover a larger number of transients than they are able to spectroscopically characterize. Photometric classification will be key to harness the full power of these surveys, both for optimizing follow-up resources and for obtaining larger and more diverse samples than those that can be spectroscopically classified.

SuperNNova is able to achieve large and high purity SN samples. We have shown that it has high performance in the classification of type Ia vs. non type Ia supernovae with accuracies up to $95\%$ for whole light-curves. When host-galaxy information is provided, accuracies increase to $98\%$. SuperNNova can be used in various classification tasks such as ternary and seven-way classification. For these classification tasks, accuracies are found to be above $90\%$ when host-galaxy information is provided. By selecting photometric samples, SuperNNova allows the exploration and characterization of diverse astrophysical objects. 

With the advent of large surveys discovering thousands of transients every night, it will be imperative to prioritize follow-up using partial light-curve classification. SuperNNova achieves unprecedented accuracy on this particularly challenging task. Our results show that we are capable of discriminating Ia and non-Ia supernova light-curves before their maximum peak with accuracies up to $86\%$ without redshift information and higher than $90\%$ when host-galaxy redshifts are provided. Classification becomes harder when classifying into many targets (3 to 7), with accuracies only reaching around $63\%$ near maximum light. For this classification task, the addition of host-galaxy redshifts can yield to accuracy improvements as high as $10$ points. Besides accuracy, inference speed is another key concern. With SuperNNova, we set a strong baseline: our algorithm is capable of classifying thousands of light-curves per second. 

Neural networks and other machine learning algorithms have been successfully applied to astrophysical analyses but their statistical properties have rarely been analyzed in detail. In this work, we devise tests to estimate the robustness of classification algorithms. We show that calibration depends on the choice of algorithm and dataset and emphasize the need for calibration diagnosis. We find that our RNNs are better calibrated than a Random Forest algorithm applied to the same dataset.

In SuperNNova, we adopt a Bayesian view of neural networks, showing much improved uncertainty estimates, calibration and out of sample behavior over standard neural networks. We have shown that the uncertainties obtained with our BRNNs reflect the confidence of predictions and exhibit the behaviour expected for epistemic uncertainties. 

One of the main goals of SuperNNova is to obtain photometrically classified type Ia supernovae. Current large surveys such as DES and PanSTARRS are already exploring methods to obtain cosmological constraints from such samples \citep{Jones:2018}. Accounting for the effect of large contamination expected in photometric samples, is one of the major limitations of these analyses. With SuperNNova we can achieve a contamination of $<2\%$, equivalent to that of current spectroscopically classified samples \citep{Rubin:2015}.

Furthermore, using photometrically classified supernovae in statistical analyses requires a thorough understanding of classification reliability and the limitations of the available training samples. We have shown that our Bayesian RNNs are reliable, well calibrated and provide meaningful uncertainty estimates. These uncertainties can be used in existing cosmology frameworks to obtain cosmological constraints \citep{Hinton:2018} or follow-up optimization frameworks to improve supernova templates \citep{Ishida:2018}. 

We have made initial assessments on the selection biases in a photometrically classified type Ia supernova sample with SuperNNova, finding a higher selection efficiency and a better population sampling when compared with spectroscopically classified supernovae in the recent DES analysis \citep{Kessler:2018}. Importantly, we have shown that out-of-distribution, random and sinusoidal shaped light-curves are mostly rejected from our photometrically classified SN Ia sample.

SuperNNova has been designed with supernova light-curve classification in mind, however it can be easily adapted to other time-domain astronomy classification problems. E.g. the classification of other transients such as kilonovae, variable stars and a wider variety of supernovae classes. We expect to continue applying SuperNNova and improving it to a wide variety of classification problems.

\section*{Acknowledgements}
We thank: Richard Scalzo and Chris Lidman for their feedback on this manuscript, Samuel Hinton for testing the framework, Rick Kessler for help with simulations and Alex Kim, Wade Blanchard and Robert Clark for useful statistics discussions.\\
Parts of this research were conducted by the Australian Research Council Centre of Excellence for All-sky Astrophysics (CAASTRO), through project number CE110001020.\\
\\




\bibliographystyle{mnras}
\bibliography{snnbib} 


\appendix

\section*{Appendix: Reproducibility} \label{appendix:reproducibility}

To validate our code, we have independently reproduced the results of \citep{Gal_a:2015,Gal_b:2015,Fortunato:2017} on a language modelling task and made the code public on GitHub. 

\section*{Appendix: Behaviour of OOD in MNIST} \label{appendix:mnist}

To better understand the behaviour of Bayesian Neural Networks, we carried out two experiments on the MNIST dataset \citep{mnist:2010}. This dataset contains images of handwritten digits.

\begin{figure}
	\includegraphics[width=1.05\columnwidth]{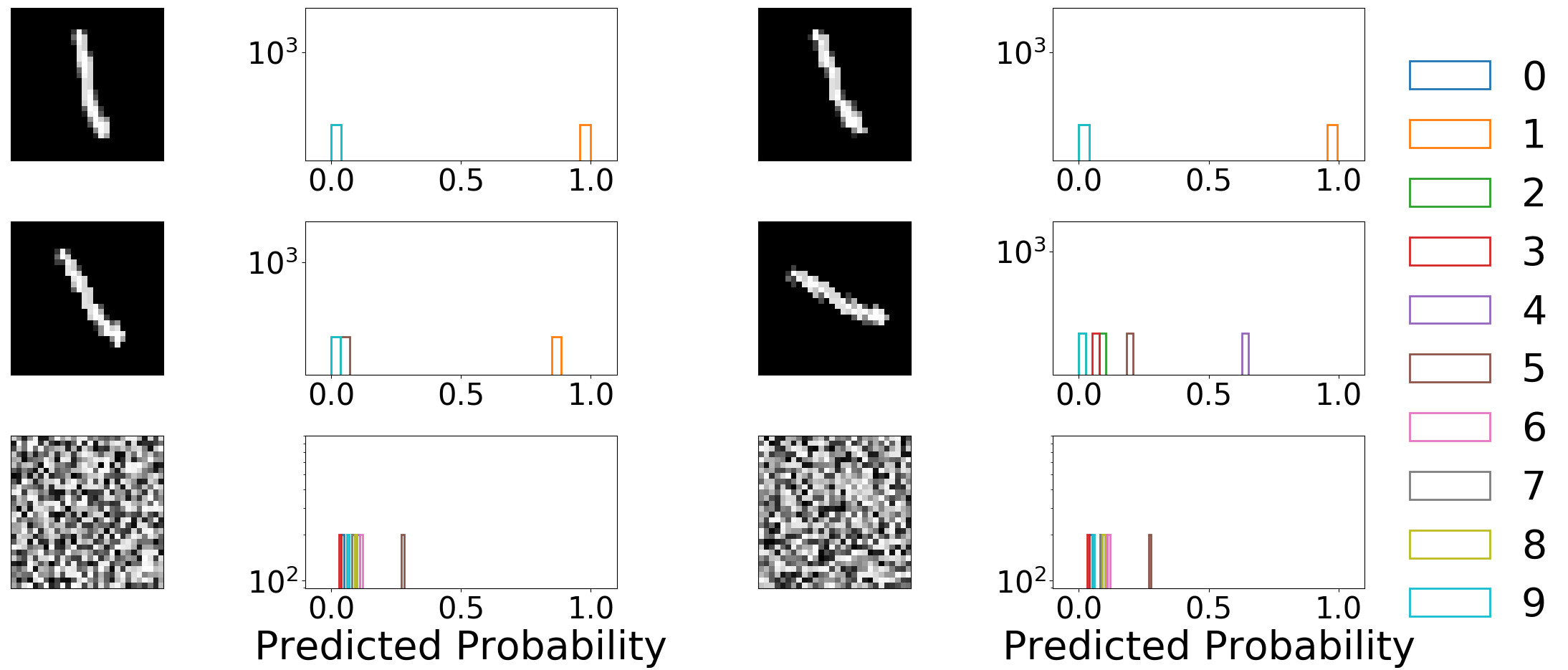}
    \caption{A standard feedforward network on MNIST. On the first and third column, images show the input given to the network. The classification probabilities for each target are given as histograms in the second and third columns.}
    \label{figure:mnist_MLP_standard}
\end{figure}

\begin{figure}
	\includegraphics[width=1.05\columnwidth]{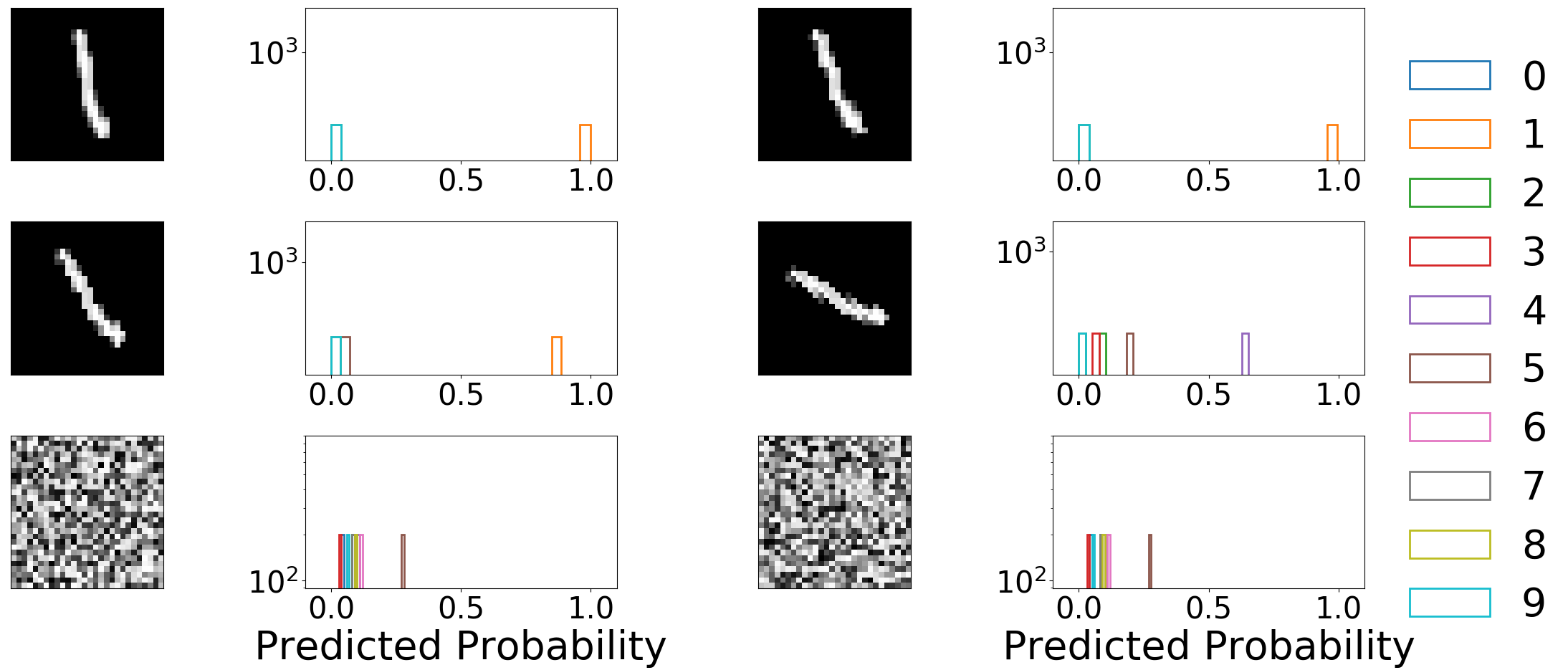}
    \caption{An MC Dropout feedforward network on MNIST. }
    \label{figure:mnist_MLP_variational}
\end{figure}

\begin{figure}
	\includegraphics[width=1.05\columnwidth]{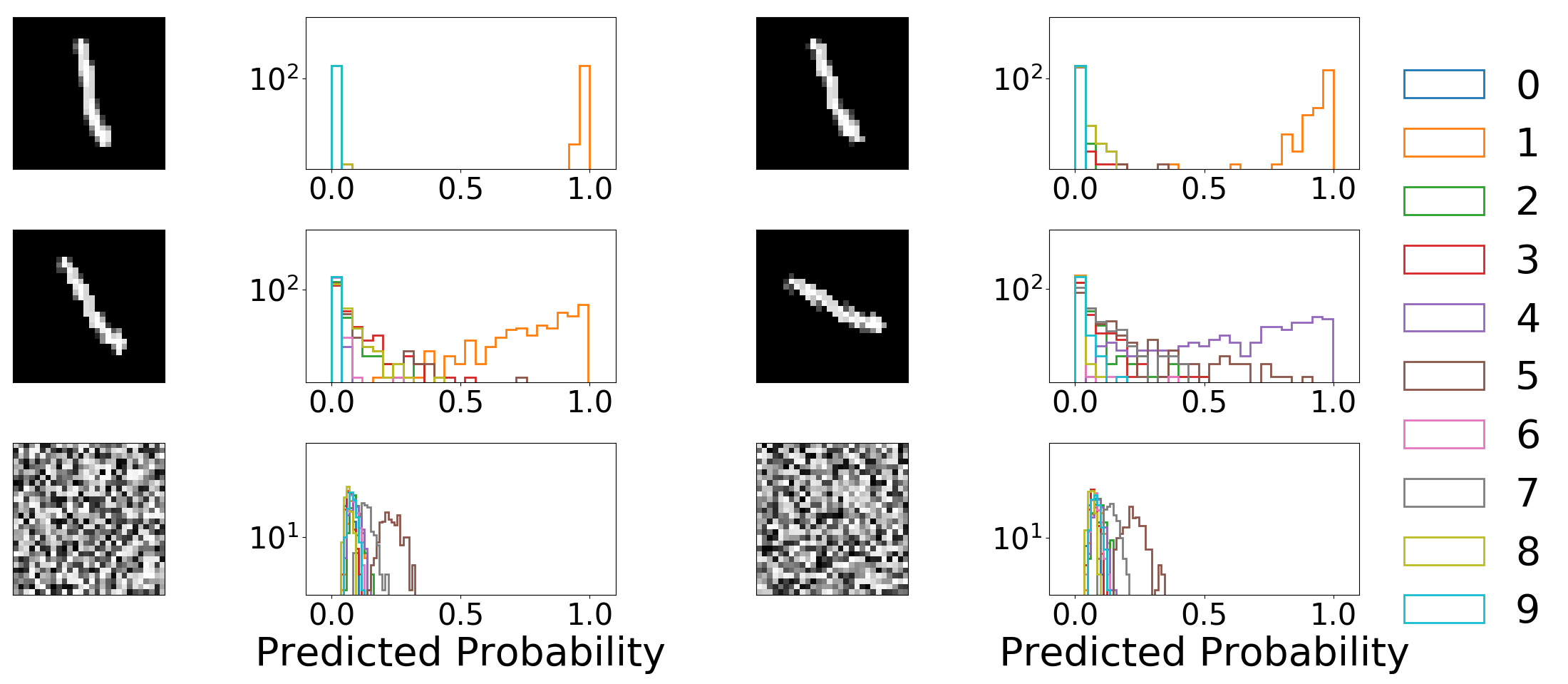}
    \caption{A Bayesian By Backprop feedforward network on MNIST.}
    \label{figure:mnist_MLP_bayes}
\end{figure}

First, we trained three (respectively standard, MC dropout and BBB) feedforward networks to classify flattened digit images into one of ten classes, each representing a digit between $0$ and $9$.  To evaluate performance on out-of-distribution data, we asked the network to carry out predictions on rotated digits, as well as completely randomized images. In Figure~\ref{figure:mnist_MLP_standard}, we show the predictions for the standard feedforward network where we see it remains confident about its prediction (the maximum of the predicted probability remains relatively high) for rotated images as well as random images. For MC dropout and BBB networks, we observe a much more interesting behaviour (Figures~\ref{figure:mnist_MLP_variational} and~\ref{figure:mnist_MLP_bayes}). When the rotation is small, the network behaves in much the same way as a non bayesian one and gives max probability with very little variance to the correct class. However, as the rotation increases, the variance of the prediction increases significantly. Clearly then, the network is uncertain about the correct class, but the fact that it still sometimes assigns a high score to one of the classes indicates that it believes the image bears some similarity to the data which it was trained on. This behaviour is in contrast with that on random images: the predicted probabilities are all concentrated to the left, with very little variance: the network is confident that it does not know how to predict this image. This is the expected behaviour for out-of-distribution samples which yields a large entropy.

\begin{figure}
	\includegraphics[width=1.05\columnwidth]{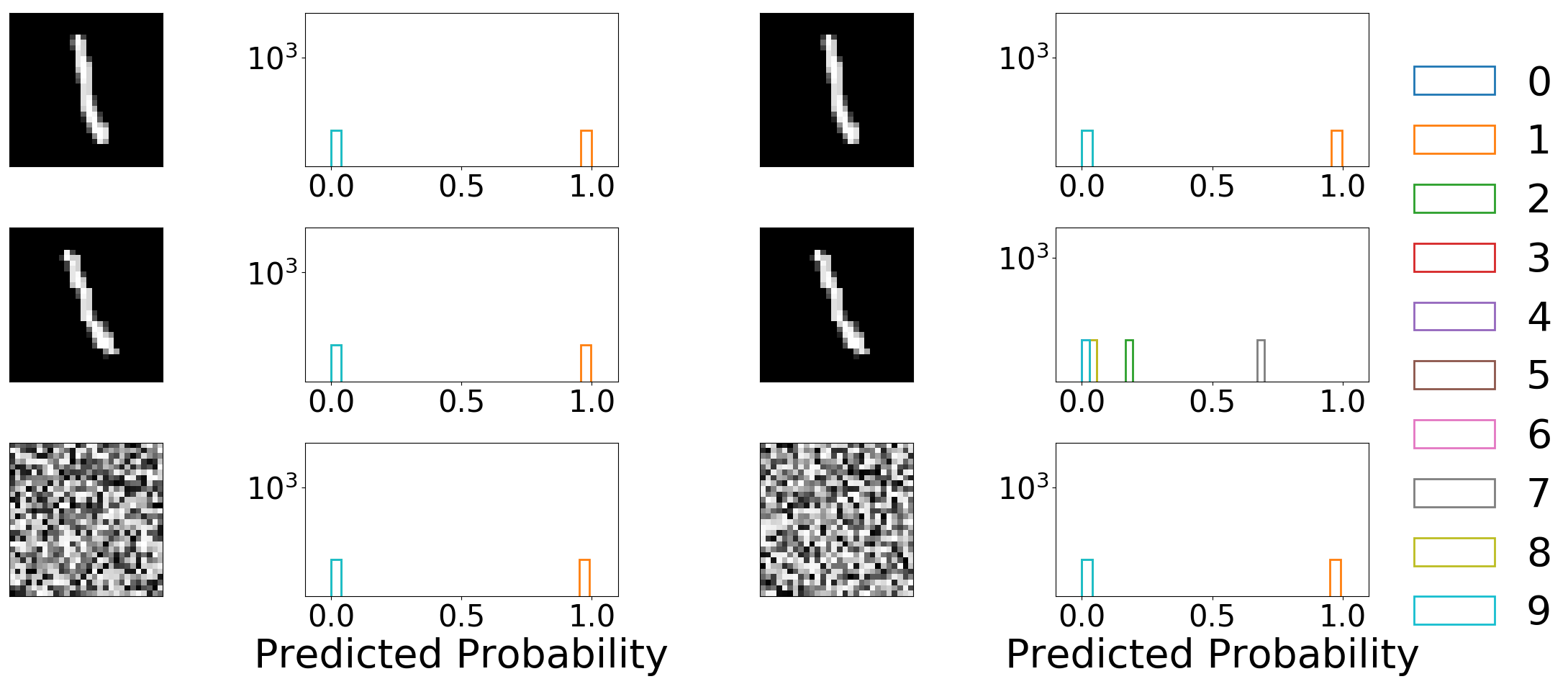}
    \caption{A recurrent network on MNIST. This RNN is able to obtain similar prediction behaviour as  Figure~\ref{figure:mnist_MLP_standard} which is what is expected for OOD events.}
    \label{figure:mnist_RNN_standard}
\end{figure}

\begin{figure}
	\includegraphics[width=1.05\columnwidth]{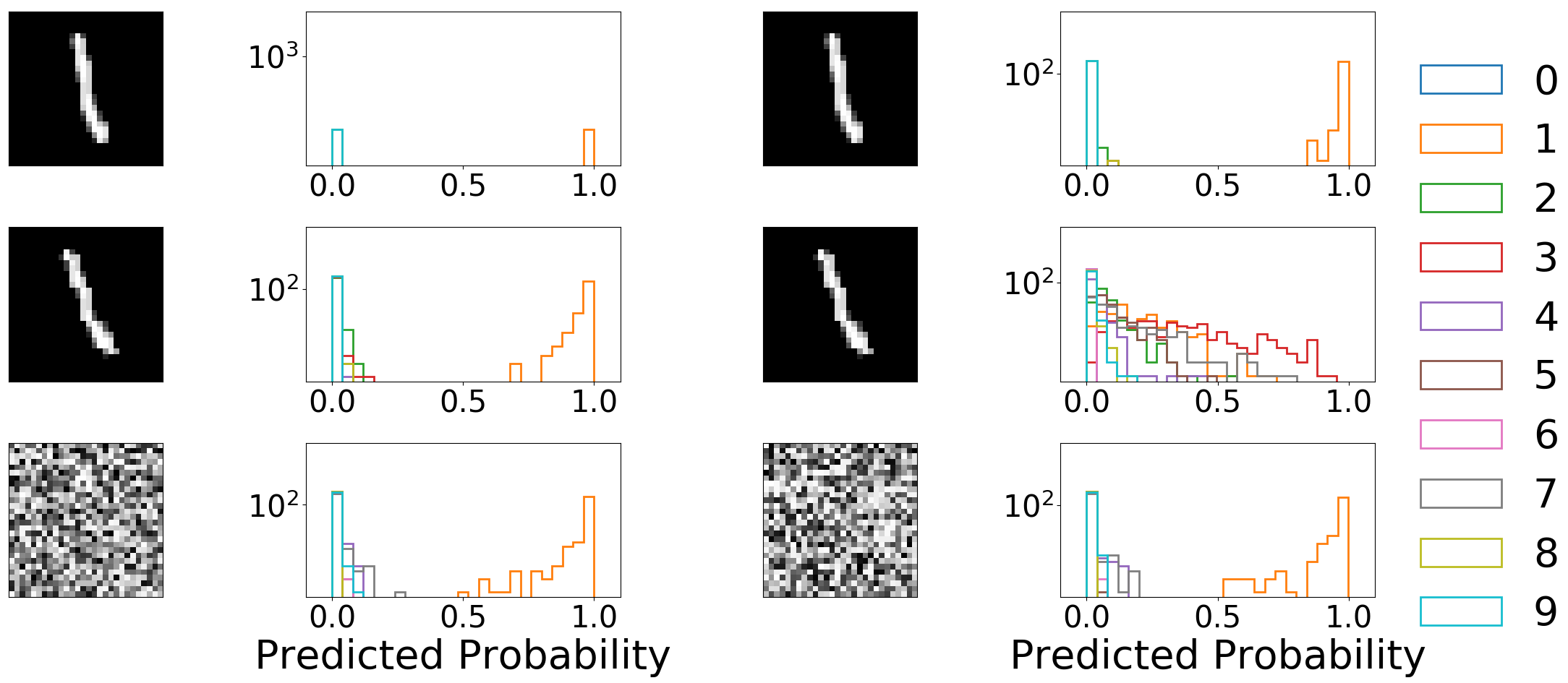}
    \caption{An MC Dropout recurrent network on MNIST. This network collapses and outputs high-certainty predictions for OOD images. }
    \label{figure:mnist_RNN_variational}
\end{figure}

\begin{figure}
	\includegraphics[width=1.05\columnwidth]{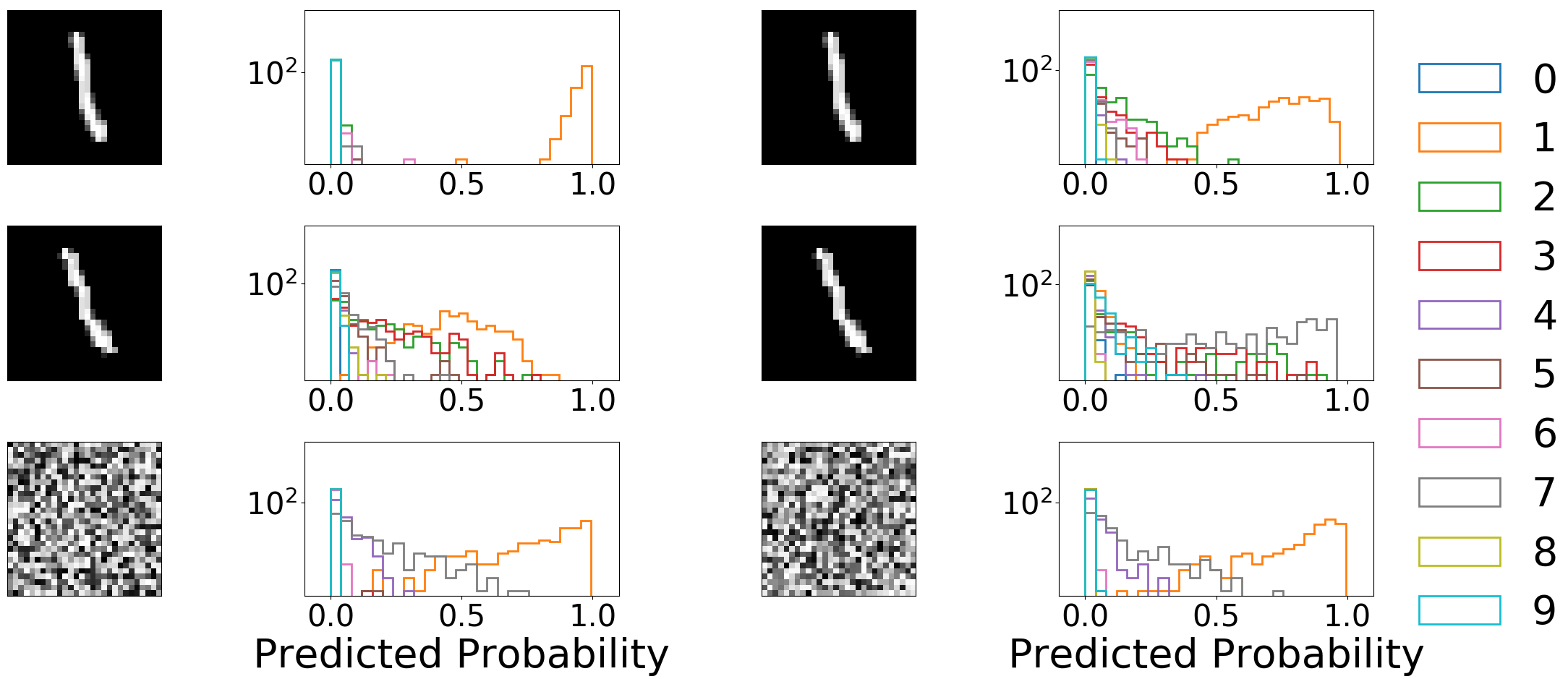}
    \caption{A Bayesian By Backprop recurrent network on MNIST. This network collapses and outputs high-certainty predictions for OOD images.}
    \label{figure:mnist_RNN_bayes}
\end{figure}

Second, we trained three recurrent neural networks (respectively standard, MC dropout and BBB), similar to the ones used in SuperNNova, to carry out the same task. The digit images were treated as a sequence, with the rows as the time steps and the columns as input features. 
The standard network behaves in much the same way as its feedforward counterpart (Figure~\ref{figure:mnist_RNN_standard}) and makes even more confident predictions on the random and rotated images. For bayesian recurrent networks, we found that it was harder to obtain desirable OOD behaviour as exhibited by the bayesian feed-forward networks, and that extensive parameter tuning was required. The MC dropout and BBB networks (Figures~\ref{figure:mnist_RNN_variational} and~\ref{figure:mnist_RNN_bayes}) show, as before, an increased variance in the prediction probabilities for random images. However, the behaviour on random data differs significantly. The MC dropout network seems to have collapsed on predicting a single class with high probability while the bayesian network exhibits large variance for multiple classes.

While we have verified that tuning the various hyper-parameters improves the uncertainty performance on this qualitative examination, it is clear that the behaviour of Bayesian recurrent networks should be critically analyzed: the network remains at risk to collapse its predictions when fed with unrelated data. This sheds light on the negative $\Delta H$ found in Section~\ref{section:ood}: for OOD data, which looks nothing like the training data, the network likely collapses and outputs a prediction with very high certainty, giving a very low entropy score to the out of sample data. We note that this is possibly exacerbated by the type of data used to train the network: supernova fluxes indeed exhibit variations spanning multiple orders of magnitude which leads to persisting artifacts even after normalization. Future work will focus on characterizing this phenomenon and developing methods to improve robustness on out-of-distribution data.

\bsp	
\label{lastpage}
\end{document}